\documentclass{amsart}
\usepackage{fullpage}
\usepackage{graphicx}
\usepackage[english]{babel}
\usepackage[latin2]{inputenc}
\usepackage[T1]{fontenc}
\usepackage{subfig}
\usepackage{bm}
\usepackage{rotating}
\usepackage{color}

\usepackage{setspace}

\theoremstyle{definition}

\theoremstyle{remark}

\numberwithin{equation}{section}
\newcommand{\x}{\mathbf{x}}

\usepackage[normalem]{ulem}%normalem<-> \emph,\em-eket nem def-ja at alahuzasra

\begin{document}

\title[Model of the hippocampal formation]{Model of the hippocampal formation explains the coexistence of grid cells and place cells}
\author[A. L{\H o}rincz et al.]{Andr\'as L{\H o}rincz$^1$, Melinda Kiszlinger$^1$ and G\'abor Szirtes$^{1,2}$}%
\address{$^1$ Department of Information Systems and
$^2$ Department of Cognitive Psychology \\
Eötvös Loránd University, Budapest H-1117, Hungary}%
\email{andras.lorincz@elte.hu}%


%\date{}%
%\dedicatory{}%
%\commby{}%
% ----------------------------------------------------------------
\begin{abstract}
In this paper we explain the strikingly  regular activity of the `grid' cells in rodent dorsal
medial entorhinal cortex (dMEC) and the spatially localized activity of the hippocampal place cells
in CA3 and CA1 by assuming that the hippocampal region is constructed to support an internal
dynamical model of the sensory information. The functioning of the different areas of the
hippocampal-entorhinal loop and their interaction are derived from a set of information theoretical
principles. We demonstrate through simple transformations of the stimulus representations that the
double form of space representation (i.e. place field and regular grid tiling) can be seen as a
computational `by-product' of the circuit. In contrast to other theoretical or computational models
we can also explain how place and grid activity may emerge at the respective areas
\emph{simultaneously}. In accord with recent views, our results point toward a close relation
between the formation of episodic memory and spatial navigation.

\end{abstract}
\maketitle
% ----------------------------------------------------------------
\section{INTRODUCTION}\label{s:intro}
%\doublespace

When we enter a new place, even without having immediately recognized each of the objects surrounding us, we need only a moment
to perceive the particular configuration of these objects within their environment (the mapping) and
define our own relative position (localization or egocentric description) in the same environment.
Sizing up distances is not of great difficulty either. In doing so we can use approximate learnt metric or
intrinsic, idiothetic (self-motion based) cues, e.g., the number of steps needed to reach the wall.
Why does spatial navigation, i.e., mapping, localization and remembering places seem so
easy for animals, whereas it still constitutes a major challenge in robotics? What are the underlying computations that provide us with a metric required to gain not only topological, but also geometrical
perception of our environment? An explanation of the surprising discovery of `grid' cells
\cite{Hafting05Microstructure} in the rodent dorsal medial entorhinal cortex (dMEC) may offer some answers to these questions.

In contrast to the spatially localized unimodal activity distribution of the place cells found
most prominently in the subfields CA3 and CA1 of the rodent hippocampus (HC)
\cite{OKeefe78thehippocampus} or, for example, in humans \cite{Ekstrom03Cellularnetworks}, the
activity of these grid cells shows more or less regular, multi-peaked activity that forms
`hexagrid' tiling of the space. Interestingly, in different layers within the dMEC, while
preserving this compact covering structure, the activity is also modulated by velocity and
directional information \cite{Sargolini06Conjunctive}. Due to this regularity, these cells are
thought to maintain a metric, and thus provide a basis for self-motion information or
`path-integration' (for a review, see \cite{McNaughton06Pathintegration}). Although this construct
is very appealing, the finding \cite{Barry07Experiencedependent} that grids may faithfully follow
the distortion of the (familiar) environment casts doubt on the straightforward link between grids
and path-integration, as such distortions may point to a topological description instead of a metric
one \cite{Dabaghian07Topological}. Acknowledging that the functional explanation of these grid
structures has yet to be found, attention has recently been focused on (1) functional links between
grid and place cells, and (2) possible mechanisms that would be able to generate such regular structures. As a
complete review is beyond the scope of this paper, here we only list some of the most recent
proposals corresponding to these two directions.

Several models in the first group elaborate on the ideas described in \cite{Sharp91Computer}:
\emph{competitive learning} resulting in sparse representations may explain the formation of place
cells in the dentate gyrus (DG),  \cite{Rolls06Entorhinalcortex} and in CA3 and CA1
\cite{Franzius07Fromgrids}. In these models the existence of an appropriately defined set of regular
grid inputs is the most stringent hypothesis. Another route is based on the ideas of
\cite{Cash99Linearsummation} on linearity and the proposals in
\cite{OKeefe05Dualphase,McNaughton06Pathintegration}: it has been shown
\cite{Solstad06Fromgridcells} that place cells can easily be formed if anatomically and
physiologically sound constraints are taken into account. The problem with this model is that it requires grids
with diverse orientations, but recent reports
\cite{Barry07Experiencedependent,Fyhn07Hippocampalremapping} show more uniformly oriented grids.

Similar ideas provide the basis for models of the second group: Linear summation of harmonic
functions forms the core idea of different oscillatory interference models
\cite{OKeefe05Dualphase,Burgess07Anoscillatory}. In this \emph{dynamic} model grid cells receive
directionally modulated oscillating dendritic inputs superimposed on somatic large scale
oscillations occurring at 4-10 Hz (theta-oscillation). With appropriate directional modulation
provided by subicular head-direction cells \cite{Ranck84Headdirection,Taube90Headdirectioncells}
this model yields regular interference patterns. To enable path-integration, grid patterns should
be precisely bound to environmental cues, because error can be accumulated in both motor signals
(speed) and direction signals. Feedback from CA1 has been suggested to provide the necessary
correction and thus to maintain the coherence of the oscillations by regulating phase resetting.
However, as CA1 is one step \emph{downstream} of the superficial layers of the entorhinal cortex
(EC), it is not obvious why it would receive at the same time a more direct sensory stimulus compared to
the information available at the entorhinal cortex. A specific class of continuous attractor models
has also been proposed either with periodic boundary conditions \cite{McNaughton06Pathintegration}
or with aperiodic boundaries, but with highly restrictive symmetric constraints on the synaptic
connection matrix
\cite{Fuhs06Aspinglassmodel}. These models achieve path-integration using the grids and can
explain many important aspects of the biological system, e.g., the similar orientation of the
grids, scaling and phase properties. However, the correct integration of signals to perform
path-integration is very sensitive only to factors related to the model setup, not to the system at hand  \cite{Burak06Doweunderstand}.

In this paper we sketch an alternative view of the problem of grid cells. Unlike the models described above, which attempt to explain a particular phenomenon or computation assigned to a given
area, we describe a functional model of the hippocampal region (HR, comprising the entorhinal
cortex, the dentate gyrus, areas CA3 and CA1, para- and presubiculum and the subiculum; see
\cite{Witter04Hippocampalformation, MohedanoMoriano07Topographicalandlaminar} in which spatial
navigation and space representation are addressed \emph{within} the more general context of efficient
memory systems. Explanation of the connections among different memory functions, such as the
formation of episodic memories, memory consolidation and retrieval, has long been recognized as one
of the major challenges in neuroscience, and several attempts have already been made to provide a
unifying view
\cite{Levy96Asequencepredicting,Recce96Memoryforplaces,Wallenstein98Thehippocampusasanassociator,Gaffan98Idiotheticinput,Redish99Beyondthecognitivemap}.
Albeit with different emphases, similar motifs emerge in most models. One such motif is that the
context for separate episodic memory traces corresponds to the environment of the actual position.
While this metaphor may help to conceptualize the acquisition of new memory traces, it does little to further our understanding of retrieval
(that is the actual usage) and consolidation   \cite{Nadel07Systemsconsolidation} of this
knowledge, as well as the role of the HR in these tasks. Here we show
that the information theoretic notion of efficient representation may link these diverse functions
and lead to a large-scale computational model of the hippocampal region in which the intriguing
grid-like activity pattern may naturally emerge. The proposed architecture is partly rooted in the
functional comparator model described in  \cite{Lorincz00Twophase,Lorincz02Mystery} and is strongly
motivated by new theoretical results on blind source separation problems
\cite{Poczos05Independentsubspace,Poczos06Noncombinatorial,Szabo07Undercomplete}.

In the Methods section, theoretical motivations about efficient representation are exposed.
Afterwards, relevant anatomical and physiological properties of the hippocampal region are
highlighted to support the resulting mapping. In the Results section (1) we formalize our model
according to the motivations described, (2) explain the functional correspondence between
the theoretical construct and the neural substrate (functional mapping) and (3) present model
verifying simulations that show how our model exhibits characteristic spatial behavior similar to that found in different parts of the HR. In the last section we discuss the relevance of our
findings, interpret our results and make predictions concerning the functioning of the HR. Finally, some
relevant but unresolved issues are enumerated.

\section{Methods}
\label{s:methods}

We begin with some definitions that we use throughout the paper.
Then we highlight the central motivations behind our large-scale
functional model.  The model is not yet extended to low-level
cellular and network mechanisms and thus the mapping of the
proposed function to the neurobiological structure is essentially
a logical arrangement of known anatomical and physiological
findings.

\section*{Theoretical motivations}
We propose a hypothesis set based on theoretical considerations. Then we enumerate the supporting
arguments for each hypothesis and explain the essential statistical concepts that form the core of
our proposal.

We use the term `memory' for internal representations of spatio-temporal patterns of observations
that in some way helps the system (agent or animal) to analyze, predict and react to changes (used
in a very broad sense). Here observation incorporates not only the perception of the external
world, but also the registering of the internal states of the self: motor commands, emotions,
goal-oriented behavior and so on. In this framework, sensory-motor binding, for example, is about to form an intermediate representation that can faithfully represent the complex observations in a
compressed form which is then used to define the response to those observations.

Motivated by ideas in machine learning, information theory and goal-oriented reinforcement
learning, one can make the following hypothesis about \emph{an efficient} memory system:

\begin{itemize}
\item \textbf{Prediction:} In order to increase the chance of survival under varying conditions,
memory creation should serve detection of novelty or change.
\item \textbf{Probabilistic interpretation:} Due to the stochastic nature of changes,
representations may only
be interpreted within a probabilistic framework.
\item \textbf{Information separation and fusion:} For tractable probabilistic inference,
the effect of the `curse of dimensionality' has to be efficiently diminished through the
discovery of the independence of the underlying causes of the changes experienced.
\end{itemize}

\section*{Prediction}

In line with \cite{Rao97Dynamicmodel,Friston05Atheoryofcorticalresponses} we hypothesize that the
goal of the memory system is to help maintain, accelerate and fine-tune a predictive coding
mechanism (for a review on predictive coding in the brain, see \cite{Kveraga07Topdownpredictions}).
The predictive faculty is needed for two reasons: not only does the agent/animal have to interact with a changeable environment, but functional delays (reaction time, internal functioning, synaptic
delays) also have to be compensated. Models of predictive coding usually employ \emph{loops} that
allow \emph{comparison} of bottom-up signals (`input') and expected signals (`output') of the
internal \emph{dynamical model} of the observations. It has already been proposed that the HR \cite{Szirtes05NeuralKalman} realizes a Kalman-filter like internal model to predict sensory
signals. Interestingly, some recent results \cite{Lorincz07Neurallyplausible} on the approximation
of \emph{independent} processes (that is dynamical models that assume independent noise as opposed
to the Gaussian noise assumption of the Kalman-filter approach) may  provide a natural combination
of efficient prediction and information extraction, thus serving both the first and the third
hypotheses.

\section*{Probabilistic interpretation}

Alternatively, the expected signals may come from a generative model
\cite{Hinton97Generativemodels} which seeks probabilistic \emph{sources} that could make up or
\emph{cause} the perceived signals: the hidden sources `explain' the observed signals. Such a
statistical approach is useful in that the system has to cope with multiple uncertainties: noisy
signals, hidden causes, faulty internal working, multiple potential interpretations. The learned
spatio-temporal structure of the hidden sources restricts the representations of the world and, in
turn, can be used for inference in a Bayesian manner  \cite{Kording04Bayesianintegration}. The
computational motivation for seeking the hidden causes is to reduce the daunting problem of
inference: the detected temporal changes are either causally related and can thus be predicted or
are intrinsically independent. If the causes are statistically independent then their joint
probability distribution may be factored.

The probabilistic framework has an added advantage compared to a deterministic encoding mechanism:
the belief of the system in its own judgment (e.g. about the existence of a particular source) may
also be explicitly encoded or maintained to support further inference \cite{Yu03Expected}.
Reconstruction networks \cite{Grossberg80Howdoes,Ullman95Sequenceseeking} try to integrate the
`best of both worlds': by maintaining an internal model of the external world, fast manipulations
of the sensory-motor integration (modulation, planning, and so on) can be achieved. On the other
hand, by extracting useful statistics of the incoming signals, robustness against noise and novelty
detection may also be realized. To the best of our knowledge, the first reconstruction network model for
brain modeling that suggested approximate pseudo-inverse computation for information processing
\emph{between} neocortical areas was published by Kawato et al., \cite{kawato93forward}. The
computational model of the neocortex was extended by Rao and Ballard
\cite{Rao97Dynamicmodel,rao99predictive}, who considered neocortical sensory processing as a
hierarchy of Kalman-filters.

The reconstruction idea has also appeared in hippocampal models \cite{Lorincz98Formingindependent}.
An extension of that model \cite{Lorincz00Twophase} suggested the integration of the early
comparator idea \cite{sokolov63perception,vinogradova75registration}. In these models, the whole
EC-HC circuitry forms a `novelty'-detecting network, in which novelty or reconstruction error is
the difference between the expected (top-down) and experienced (bottom-up) neuronal
representations. The proposed model successfully predicted independence in the cellular activity in
CA1 \cite{Redish01Independenceoffiring} and was the first to suggest distinct roles for the direct and
tri-synaptic pathways  \cite{Kloosterman04Tworeentrant}. A reconstruction-network like mechanism\cite{Hasselmo02Aproposedfunction} connecting CA3 and CA1 has been suggested that directs the
information flow during encoding and retrieval. In another study \cite{becker05computational}, each
hippocampal layer forms a separate representation that could be transformed linearly to reconstruct
the original activation patterns in the EC.

These lines of arguments lead to the first assumption about functional mapping: the HR may be
considered as a \emph{reconstruction-network} with \emph{predictive capacity}.

\section*{Information separation and information fusion}\label{sss:infoseparation}

For any probabilistic reasoning, we have to define the elementary events that make up all possible
outcomes. Without knowing their true probability distribution, we need to sample them (by
experiencing different outcomes) and approximate the unknown distribution. This task becomes
computationally intractable with the increasing number of possible events. Furthermore,
discretization of the space-time continuum, e.g., sampling is another source of noise and
computational explosion.

Consider, for example, the problem of sequence learning \cite{Fusi07Aneuralcircuitmodel}. If we
want to take into account all pieces of sensory information at each moment  during which the system is able to take a
sample, we can only store sequences of limited temporal duration. Furthermore, the number
of patterns that make up the sequence is not known \emph{beforehand}. In turn, the system should be
able to flexibly compress the  spatiotemporal patterns into an internal form which (1) is subject
to memory capacity constraints, but (2) still preserves all relevant information concerning the ongoing
events. To do so, information should be collected, represented, and possibly compressed over time,
because (spatial) changes take place at different \emph{temporal} scales compared to the internal
clock. Motion induced visual changes, for example, imply that part of the information is lost
unless it is remembered in some economical forms.

Temporal compression can be achieved by implementing a predictive system which can recover (explain
in simpler terms) the deterministic parts of a stochastic process. If the predictable part is extracted, the rest of the available information (the so called `innovation') has reduced temporal
correlation.

On the other hand, if the independence of the underlying causes may be assumed (as noted above), information transfer can be optimized by forcing independence among the components of the emerging representation
\cite{Jutten91Blindseparation,Comon94Independentcomponent,cichocki94robust,laheld94adaptive,bell95information,amari96new}.

Importantly, the very same assumption may greatly simplify the predictive modeling as well. This is
in line with Barlow's revised formulation of the redundancy reduction principle
\cite{Barlow01Redundancyreductionrevisited}: representations should not be rigid structures but rather
tools that serve the animal's current (that is variable) goals. They should therefore appropriately map
the changing statistics of the world they represent.

Elaborating on his idea about obvious (simple) and `hidden' forms of redundancy (see
\cite{Barlow01Redundancyreductionrevisited} and the references therein), the second main functional
conjecture in our model is that HR maximizes information transfer throughout the neural circuitry
(by reducing the obvious redundancy) \emph{and at the same time} reveals the hidden structures by
separating them into independent subspaces. That is, the learning system reveals the types of
approximately independent sources and their own intrinsic dimensionality.

To highlight this issue, consider the problem of space representation formed by the HR. The main
input through MEC to the HR is primarily multimodal sensory information with implicit and limited
spatial information content \cite{Fyhn04Spatialrepresentation}, such as direction or configuration.
Configuration of objects can be interpreted as one of the independent descriptors of the
environment. However, its true or approximate dimension can only be revealed if the system is able
to detach the corresponding correlations among the components of the representation from those that
carry information about other physical aspects, such as texture or color. Such separation may reveal
that configuration may best be described in a 2 or 3 dimensional space that actually corresponds to
our abstract notion of Euclidean space \cite{Dabaghian07Topological,Dabaghian07Topologicalmaps}.

Dimensionality, in general, may not be well defined for the other physical aspects, such as texture
or color, see, e.g., \cite{ben-shahar04geometrical}. Note that these descriptors, or `factors'
assume each other, but they are also highly independent. This dichotomy can be exploited in the
following way. On the one hand, there is combinatorial gain in the description of events if
characterization, categorization and prediction of the factors takes place separately. For
instance, a screenshot of an animal is a static image containing no direct information concerning motion. Still, the particular combination of
factors or components of the animal may help to draw inferences concerning the unseen parts of the animal
and the (intended) direction of motion. \emph{Pattern completion} can be seen as a particular
inference problem that occurs in space and time.

Interestingly, as new results \cite{Poczos05Independentsubspace} on blind source separation show,
factorial coding and subspace separation can be achieved \emph{simultaneously}. In blind source
separation problems, not only the sources, but also the mixing process that generates the received
signals are unknown. In general, this problem cannot be solved without regularization. Assuming
independence in time seems plausible in many problems. For a special case of \emph{instantaneous}
linear mixtures of (statistically) independent and identically distributed (i.i.d), one dimensional
sources, where the dimension of the signal is larger than or equal to the dimension of the
sources, there exist efficient, neurally plausible Independent Component Algorithm (ICA) algorithms
\cite{Giannakopoulos98experimental,linksker99local} that can recover the true non-Gaussian lower
dimensional sources by \emph{demixing} the signal.

ICA can be significantly faster \cite{amari96new} if separation
is preceded by \emph{whitening}. This intermediate transformation
reduces the instantaneous (zero time lag or \emph{spatial})
second-order correlations (i.e., it decorrelates) and it also
normalizes the signals. Informally, decorrelation transforms the
data onto an orthogonal subspace such that the projection of the data onto the first (principal) direction of the subspace has the greatest variance, projection on the second principal direction has the second greatest variance and so on. The decorrelation
part is also called Principal Component Analysis (PCA) and may be
used for dimension reduction in an informed way as it provides a
measure of how much information (at the second-order correlation
level) is lost by ignoring the last $k$ directions or components.
Whitening admits that all sources may equally be important, so
after the decorrelation step it equalizes the variances of the
components. The terms `whitening' and `decorrelation' may be used
interchangeably, but they scale the results differently.

For more general cases of ICA, there is no trivial solution, but as both experience and
\cite{cardoso98multidimensional,Poczos05Independentsubspace} several theoretical advances have indicated \cite{Szabo07Undercomplete,poczos07independent}, sources can in many cases be recovered
even if conditions (independence, i.i.d. properties or equal dimensions) are not met. The recovered
components can be \emph{grouped} by their mutual information --- that is using the
`non-independence' information --- thus revealing the number of separable sources and the
dimensions of their subspaces. This procedure factorizes the information and gives rise to
combinatorial gains in the \emph{storage} requirements. In addition, recent theoretical findings
allege that the \emph{search} for these factors can be accelerated in a non-combinatorial way
\cite{Poczos06Noncombinatorial,Lorincz07Neurallyplausible,szabo08autoregressive} even if the
dimensions of the subspaces are not known beforehand  \cite{poczos07independent}.

\section{Known anatomical and physiological constraints}\label{ss:HR}

\begin{figure}%
\includegraphics[angle=-90,width=14cm]{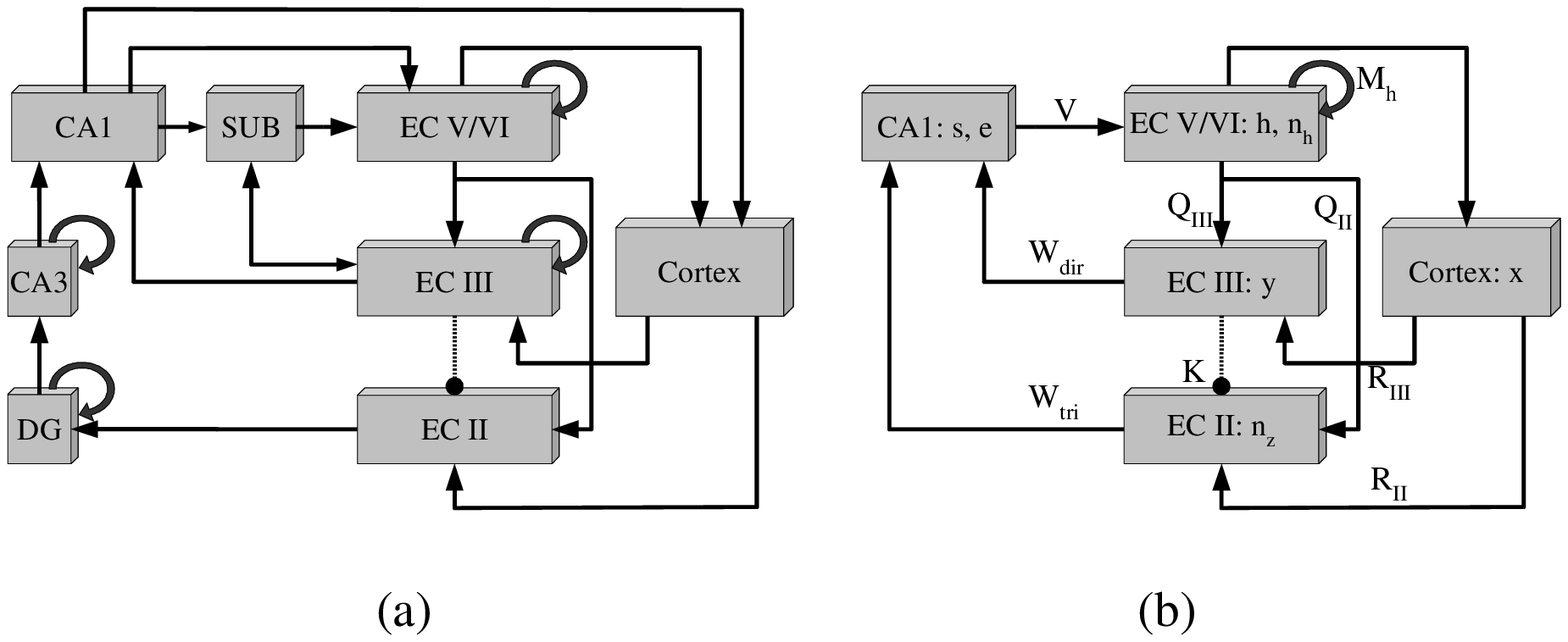}%
\caption{(a): Diagram of the main connections of HR. Arrows denote excitatory connections and solid
circles denote mostly inhibitory connections. (b): Connections playing a role in the model. Roman
letters denote the layers in the entorhinal cortex (EC), \newline $\mathbf{x}$: signal from cortex,
\newline $\mathbf{y}$:   whitened input at EC III,
\newline $\mathbf{n_z}$: whitened novelty (or innovation) of the input at EC II,
\newline $\mathbf{h}$: hidden model at EC deep layers,
\newline $\mathbf{n}_{\mathbf{h}}$: innovation of the hidden model at EC deep layers,
\newline $\mathbf{s}$: ICA output at CA1 during positive theta phase,
\newline $\mathbf{e}$: ICA output at CA1 during negative theta phase,
\newline $R_{II}$ and $R_{III}$: postrhinal to EC II and postrhinal to EC III efferents, respectively,
\newline $Q_{II}$ and $Q_{III}$: EC deep layers to EC II and EC III connections, respectively,
\newline $K$ inhibitory feedback from EC III to EC II.
\newline $V$: CA1 to EC deep layer efferents, \newline $M_h$: recurrent collaterals
at the deep layers of the EC, \newline $W_{tri}$ and $W_{dir}$: tri-synaptic and direct connections
between EC superficial layers and the CA1 subfield, respectively.}\label{fig:struct}
\end{figure}

In this section, we describe those characteristics of the HR that guide and constrain our model.
The circuitry of HR (left panel of Fig.~\ref{fig:struct}) has several unique properties that probably
contribute to its central role in all memory functions. Here we highlight features that seem
relevant for mapping the functions onto the neural substrate.

\section*{Direction of information flow}\label{sss:dirflow}

First, there is a dominantly unidirectional (\cite{Naber01Reciprocal}, but see
\cite{Shao05Electrophysiologicalevidenceusing}), and parallel connection system among all parts:
superficial layers of EC receive input from adjacent cortical regions and transmit the signals
toward CA1 and the subiculum mediated by CA3. This transmission, however, is not a simple relay: it
takes place in a tightly controlled way using two separate routes:  the so-called
\emph{tri-synaptic} connection system (EC II--DG--CA3--CA1) and the \emph{direct} route from EC III
to CA1. As the exact nature of the input received by EC II and EC III is not known and we want to
focus on the functioning \emph{within} the HR, we assume that the superficial layers share the same
cortical input. We also assume that differences in the activity of these layers stem from their
differing intrinsic physiology (e.g. the ratio of interneurons that enables strong feedforward
inhibition in EC II), anatomy (role of recurrent collaterals) and the received feedback (EC layers
V/VI project back to both layers and  EC III receives signals from the subiculum, too).

CA1 and the subiculum, which are considered to be the  main output regions of the HR, project back
to the deep layers of EC. In parallel with the subicular pathway, CA1 is linked to the deep layers
directly as well. The parallel systems in part preserve topographical arrangement
\cite{Witter06Connectionsofthesubiculum} but there exists a separation along the lateral to medial
direction. The lateral and medial parts of the entorhinal cortex (LEC and MEC, respectively)
receive input from different cortical areas and, in turn, project to non-overlapping portions of
CA1 and the subiculum. In contrast, DG and CA3 receive convergent input from both LEC and MEC.
An important functional consequence is that the fusion of spatial and non-spatial information may
be strictly controlled within HR
\cite{Gigg06Constraintsonhippocampalprocessing,Witter06Spatialrepresentation}.

The EC deep layers, which presumably also receive modulatory or control signals from different
cortical areas, close the loop: they send mostly excitatory  \cite{vanHaeften03Morphological}
feedback to the superficial layers.

\section*{Unique intra-regional interactions in each area}

Although place cells can be found everywhere in DG, CA3 and CA1, their coding mechanism may be
quite different, as the underlying connection systems have significantly distinct features. DG is
unique for its temporally tunable connections \cite{Henze02single}. CA3 has a dense collateral
system which has a particular role in memory replay
\cite{Louie01Temporallystructured,Foster06Reversereplay,diba07forward,csicsvari07place,oneill07reactivation}.
CA1, as a single exception in the whole circuitry, has no recurrent collaterals and the activity of
the principal cells seems to be independent  \cite{Redish01Independenceoffiring}.

\section*{Temporal synchrony across and within different areas}

In addition to the intricate anatomy, the physiology of the separate modules is also striking. The most
prominent feature is the interplay between different forms of oscillatory activities, the
synchronized membrane potential oscillation between the 4-10 Hz theta and the 40-100 Hz gamma
frequency bands,  \cite{Bragin95Gammaoscillation, Canolty06Highgammapower}, which have differential
effects on the different modules. Several functional roles have already been assigned to these
activity forms , such as the control of synchrony throughout the circuitry
\cite{Denham00ModelofThetarhythm} or the provision of an internal reference clock
\cite{jeffreys96neuronal,jensen96physiologically}.

The main generator of theta is thought to be in the septum (which is the only extra-hippocampal
target of CA3), but layer EC II may also be able to initiate theta activity. The reciprocity
between the subiculum and the HR via CA3 may suggest that HR has a sophisticated mechanism for
self-regulating synchrony. In addition, EC II neurons are theta modulated and show phase
precession, similarly to the place cells in the hippocampus
\cite{Hafting05Topographicorganization}.

EC III, which is very close to layer EC II, however, is phase locked to the main theta and can
maintain persistent activity  \cite{Tahvildari07Switching}. Deep layers of the EC show peculiar
functioning as well. In contrast to the superficial layers, EC V can generate input specific graded
persistent activity in individual neurons \cite{Egorov02Gradedpersistentactivity} which is
generally considered the underlying neural mechanism of working memory
\cite{GoldmanRakic95Cellularbasis}. Furthermore, the relative homogeneity of the CA1 response to
changing inputs as compared to that seen in the deep EC may suggest
\cite{Frank06Hippocampalandcorticalplacecell} that active CA1 neurons are engaged in representing
one environment, while deep EC may contain multiple subpopulations, some tied to CA1 output while
others are more independent of CA1. Interestingly, separate modules or `cell islands' can be found
in EC II as well \cite{Witter06Spatialrepresentation}. As a consequence, if deep layers can
represent several likely models concerning the world, there should be a switching mechanism that can
help select the one that best serves correct predictive coding. It is intriguing that layer III of
the EC has been found to receive such switching signals  \cite{Tahvildari07Switching}. Last, but
not least, signals carrying different aspects of spatial information, such as position,
head-direction or speed, seem to interfere at several stages. While activity in CA3 and CA1
doesn't show correlation with directional information, postsubicular head-direction cells directly
innervate the deep layers of EC, which in turn send this information to the superficial layers.
According to this scenario, grid cells in EC III show clear conjunctive correlation representing
mixed information at the same time \cite{Sargolini06Conjunctive}. However, the activity of neurons
in EC layer II is free of directional modulation.

\section{Results}\label{s:results}

In the first part we formalize the proposed functions by providing a mathematical construction. In
the resulting computational model the different functional modules are not yet anchored to the real
system. Since this reverse-engineering approach (assignment of the functions precedes the
description of the structure) is essentially ill-posed (offering several solutions), in the second
part we attempt to map the modules onto the real neural system by taking into account the
biological constraints collected in the previous section. Finally, simulations are presented in
which the function of the model is demonstrated on inputs that can be related to signals received by
the hippocampal region.

\section*{Results I: Formal description of the functional model}\label{s:resultsI}

Let us assume the system's goal is to  form efficient
representation of the sensory information which can be used for
prediction. Efficiency refers to storage capacity (a small number of
`factors' should be used to reconstruct large number of possible
inputs) and speed (the system should  try out only a few
combinations of the factors). Prediction is the ability to
generate expected inputs. Let us begin with an abstract
description of the observation of the external world (At this
point we don't model different sensory modalities. The input
variable is simply a description of the external world). The
sensory input $\mathbf{x}(t)$ to the system may be assumed to be a
mixture of hidden source signals or causes:

\begin{equation}\label{e:ICA_input}
\mathbf{x}(t) = A\mathbf{s}(t),
\end{equation}
where $A\in \mathbb{R}^{n\times n}$ is a mixing matrix, and
$\mathbf{s}(t)\in \mathbb{R}^{n}$ are the sources to extract.
Regarding our hypothesis (3), ICA is designed to solve a similar
problem under the condition that the components of $\mathbf{s}$
are i.i.d., and statistically independent. However, the observed
quantities may not be i.i.d.,
\begin{equation}\label{e:hidden_process_input}
\mathbf{s}(t+1) = F\mathbf{s}(t)+\mathbf{e}(t+1),
\end{equation}
where $\mathbf{e}(t)$ is called the `driving noise', `true source', or `innovation'. The expression
`driving noise' refers to the fact that process $\mathbf{s}$ is maintained by the `true source'
$\mathbf{e}$: without this input,  $\mathbf{s}(t)$ would decay. Due to the mixing effect of matrix
$F$ which describes the deterministic part of the process, the components in $\mathbf{s}(t)$ are not
independent anymore. Obviously one can envision more sophisticated systems. Nevertheless, for
higher order processes or signals with echoes, the formalism can be brought to very similar forms
\cite{Lorincz00Twophase,Szabo07Undercomplete,poczos07independent}. As long as the components of the
true source, $\mathbf{e}(t)$ can be considered  independent, the efficient representation can again
be achieved by extracting these components. If the dynamics are `weak' in the sense that only weak
temporal correlations are introduced by $F$, then we arrive at the original ICA problem. Because we
are interested in the causes, i.e., in the driving noise, we need to learn both the autoregressive
process ($F$) and the mixing process ($A$). This can be achieved
\cite{Lorincz07Neurallyplausible} only if components of the true driving noises are independent. Under
the normal (Gaussian) noise assumption the effects of these processes cannot be distinguished. We
need to carry out some manipulations in order not to misguide ICA.

We make use of the identities
\begin{equation}\label{e:ICA_input_1}
\mathbf{x}(t+1)=A\mathbf{s}(t+1)=AF\mathbf{s}(t)+A\mathbf{e}(t+1))
\end{equation}
to get
\begin{equation}\label{e:ICA_input_2}
\mathbf{x}(t+1) = M\mathbf{x}(t)+\mathbf{n}(t+1),
\end{equation}
where $\mathbf{n}(t+1)\doteq A\mathbf{e}(t+1)$ and $M\doteq
AFA^{-1}$ under the assumption that matrix $A$ can be inverted.
Thus, both Eq.~\eqref{e:hidden_process_input} and
Eq.~\eqref{e:ICA_input_2} have autoregressive forms. Due to the
mixing effect of $A$ (Central Limit Theorem), the distribution of
$A\mathbf{e}(t+1)$ is more Gaussian-like compared to the true
sources. It implies that the standard solution of the Gaussian
autoregressive processes can be applied as the first step to
unfold the hidden processes.

Now let us suppose we have a tunable system and our task is to find the hidden
process $\mathbf{s}$ and the driving source $\mathbf{e}$ using only the observation  $\mathbf{x}(t)$. In what
follows, we distinguish approximations of the true quantities by a
small \emph{hat}.

First, one can remove the autoregressive part by estimating matrix
$\hat{M}$ through the minimization of the following cost function
\begin{equation}\label{e:AR_M_cost}
J(\hat{M}) =  \frac{1}{2} \sum_t |\mathbf{x}(t+1) - \hat{M}(t)\, \mathbf{x}(t)|^2.
\end{equation}
for all available data pairs $(\mathbf{x}(t+1),\mathbf{x}(t))$.
Then, we have a model that predicts the next expected input
\begin{equation}\label{e:AR_M_est}
\mathbf{\hat{x}}(t+1) =  \hat{M}(t) \mathbf{x}(t)
\end{equation}
and we can estimate the innovation, i.e.,  the difference between
the observed input and the expected input at time $t$:
\begin{equation}\label{e:AR_x_inno}
\mathbf{\hat{n}_x}(t) = \mathbf{x}(t) - \mathbf{\hat{x}}(t).
\end{equation}
For Gaussian $\mathbf{\hat{n}_x}(t)$, the minimization of
Eq.~\eqref{e:AR_M_cost} leads to the following gradient rule:
\begin{equation}\label{e:ARlearn}
\Delta  \hat{M}(t+1) =  \alpha_t \, \big(\mathbf{x}(t+1) - \hat{M}(t) \, \mathbf{x}(t)\big) \mathbf{x}(t)'
=\alpha_t \,\mathbf{\hat{n}_x}(t+1)\mathbf{x}(t)'
\end{equation}
where prime $'$ denotes the transposed form for vectors and also for matrices, and $\alpha_t$ is
the learning rate. If $\alpha_t$ diminishes according to some suitable schedule then $\hat{M}(t)$
converges to the real $M$ \cite{Robbins51Stochastic}. In what follows, the learning rules will be
written as
\begin{equation}\label{e:ARlearn_}
\Delta  \hat{M}(t+1) \propto \, \mathbf{\hat{n}_x}(t+1)\mathbf{x}(t)'
\end{equation}
where the sign `$\propto$' denotes the Robbins-Monro schedule.
Note, however, that if the world is changing then it is  better to
maintain adaptation forever.

So far we have exploited the Gaussianity property of the driving noise
to learn the dynamical system. Now we can make use of the fact that
upon convergence, the innovation term also converges to the mixed
true sources of Eq.~\eqref{e:ICA_input} ($\mathbf{n_x}(t)\mapsto
A\mathbf{e}(t)$). In turn, simple separation of the innovation
yields the demixing process $W$, which is the approximation of the
inverse of the mixing matrix: $W=\hat{A}^{-1}$. Then
$\mathbf{\hat{e}}(t)=W\mathbf{\hat{n}_x}(t)$ is the approximation
of true sources, whereas $\mathbf{\hat{s}}(t)=W\mathbf{x}(t)$
approximates the hidden process.

One can approximate the autoregressive matrix $F$ using quantities $\mathbf{x}$, $\hat{M}$, and
$\mathbf{\hat{e}}$. The goal of the approximation is to optimize prediction, that is, to minimize
the following cost function:
\begin{equation}\label{e:AR_F_cost}
J(\hat{F}) =  \frac{1}{2} \sum_t |\mathbf{\hat{s}}(t+1) - \hat{F}(t)\, \mathbf{\hat{s}}(t)|^2.
\end{equation}
As with matrix $M$, matrix $F$ can be learned through the following gradient rule:
%\begin{equation}\label{e:hiddenARlearn_a}
%    \Delta  \hat{F}(t+1)  \propto q, \beta_t \, \big(\mathbf{\hat{s}}(t+1) - \hat{F}(t) \, \mathbf{\hat{s}}(t)\big) %\mathbf{\hat{s}}(t)'
%    =\beta_t \,\mathbf{\hat{e}}(t+1)\mathbf{\hat{s}}(t)',
%\end{equation}

\begin{equation}\label{e:hiddenARlearn_a}
\Delta  \hat{F}(t+1)  \propto \,   \big(\mathbf{\hat{s}}(t+1) - \hat{F}(t) \, \mathbf{\hat{s}}(t)\big) \mathbf{\hat{s}}(t)'
= \,\mathbf{\hat{e}}(t+1)\mathbf{\hat{s}}(t)',
\end{equation}
that is,
\begin{equation}\label{e:hiddenARlearn_b}
\Delta  \hat{F}(t+1)  \propto \,    \,W\mathbf{n_x}(t+1)(W\mathbf{x}(t))'
\end{equation}
This strategy has been detailed in
\cite{Lorincz07Neurallyplausible}.

Let us note that the gradient learning rules of Eqs.~\eqref{e:ARlearn} and
\eqref{e:hiddenARlearn_a} may have plausible neural implementations as they are incremental and the
change in one synapse does not depend on the change in all the other synapses. If this latter
condition is met, then we say that learning is \emph{Hebbian}, or alternatively, the learning rule
is \emph{`local'}.

As signals should be separated and --- as was argued before
--- separation  can be facilitated if whitening
takes place first, a decorrelation stage might be introduced. According to
\cite{Cardoso96Equivariantadaptive}, signals
$\mathbf{y}=P_{\mathbf{y}}\mathbf{x}$ become decorrelated if
\begin{equation}\label{e:white_x}
\Delta P_{\mathbf{y}}(t+1)' \propto P_{\mathbf{y}}(t)'(I-\mathbf{y}(t)\mathbf{y}(t)')
\end{equation}
for all times $t=1,2,\ldots$ and under suitable conditions. Note that here, in
Eq.~\eqref{e:white_x}, and in similar equations later, the learning rule contains the transposed
form of matrix $P_{\mathbf{y}}$ and thus dimension reduction ($\dim(\mathbf{y}) \le
\dim(\mathbf{x})$) is possible. Intuitively this serial update algorithm pushes the covariance
matrix of $\mathbf{y}(t)$ ($E(\mathbf{y\,y}')$, where $E(.)$ denotes expectation) to become
identity. Let us remark that there are many artificial neuronal implementations of such algorithms
\cite{foldiak90forming,hyvarinen98independent,linksker99local,basalyga03statistical}.

For the very same reason, innovation $\mathbf{n_x}(t)$ should also be decorrelated. The linear transformation
$\mathbf{n_z}=P_{\mathbf{n_z}}\mathbf{n_x}(t)$ of innovation
$\mathbf{n_x}(t)$ becomes white if tuning of $P_{\mathbf{n_z}}$ is
as follows:

\begin{equation}\label{e:white_n}
\Delta P_{\mathbf{n_z}}(t+1)' \propto P_{\mathbf{n_z}}'(t)(I-\mathbf{n_z}(t)\mathbf{n_z}(t)')
\end{equation}
at time $t$.

Statistically independent sources from $\mathbf{n_z}$ can be extracted via a nonlinear modification
\cite{Cardoso96Equivariantadaptive} of update rule Eq.~\eqref{e:white_n}. There are many variants
for this non-linear learning rule and we provide the simplest of these here:
\begin{equation}\label{e:separation}
\Delta W_{\mathbf{n_z}}(t+1)' \propto W_{\mathbf{n_z}}(t)'(I-\mathbf{\hat{e}}(t)\,
f(\mathbf{\hat{e}}(t))').
\end{equation}
Here, $f(\cdot)$ is an (almost) arbitrary component-wise nonlinear function. Upon convergence,
the components of $\mathbf{\hat{e}}(t)=W_{\mathbf{n_z}}\mathbf{n_z}(t)$ approximate the components
of the independent source $\mathbf{e}(t)$ apart from an arbitrary permutation in the order of the
components, their scale and sign. Interestingly, spike timing dependent plasticity has been
suggested to realize this non-linear learning rule \cite{NIPS2005_533}.

The learning equations of the whitening and separation processes have several
implications concerning possible mappings.
\begin{description}
\item[Two stages] Removal of the temporal correlations precedes the extraction of the independent factors.
\item[Two channels] According to Eq.~\eqref{e:hiddenARlearn_a}, the process of learning the predictive system
requires concurrent access to the input and the innovation. These variables may be stored
separately and conveyed to the predictive layer via separate channels.
\item[Identical separation]  It can be seen from Eq.~\eqref{e:hiddenARlearn_b}
that both $\mathbf{\hat{e}}(t)$ and $\mathbf{\hat{s}}(t)$ are demixed by the same matrix, so
they should be processed in the same demixing channel (violating the conjecture above)
or there should be a mechanism that can compensate for the differences (e.g., sign and permutation of the
components) in the linear transformations \emph{in two channels} for proper demixing.
\end{description}

\section*{Results II: Functional mapping of the model}\label{s:resultsII}

Since both the computational considerations and the anatomical findings are quite complex, we  need to introduce some simplifications:

\begin{description}
\item[Rate coding] How information is actually transmitted by the neurons is neglected. The key issue is that
once the particular form is given, the function of the system can be analyzed as an information processing system.
(On the controversies concerning the potential forms of information processing, however, see
e.g. ~\cite{Reyes03Synchronydependent,Masuda07Dualcoding}.) Our system description becomes
simpler if we use analog values, which corresponds to the concept of rate coding as opposed to spike based
temporal coding. The supposed low-pass filtering effect of the theta oscillation also suggests
that for some functions fine scale temporal precision might be neglected.
\item[Laminar homogeneity] We neglect the complexity and richness at the cellular level and consider neurons
as computational units. The computations may change from layer to layer, but within a layer the nature of the computation is the same
for all neurons. This corresponds to the terminology of standard artificial neural networks.
\item[Apparent linearity] Although strong nonlinearities are present everywhere,
from the subcellular level to the network level, there are nonetheless many cases in which the overall response of the
system is approximately linear, see, e.g., ~\cite{linksker99local},
\cite{hsu04quantifying}  and \cite{escabi05contribution} and the cited references. The complex contrast normalization mechanisms in visual sensory processing  may constitute a specific example \cite{Finn07Theemergenceofcontrastinvariant}.
\end{description}

>From now on, matrices denote synaptic weights (connection
strength) between layers and vector denotes the activity at a given
layer. We shall slightly abuse notation and will discard the hats
from our equations, as all learned quantities are approximations.

Figure~\ref{fig:struct} may help to understand the modular
structure of our model and its relation to the hippocampal region.
While the left panel depicts the gross anatomy of the areas,
including the different connection systems, the right panel of
Fig.~\ref{fig:struct} shows the simplified architecture and the
functional correspondences.

The following areas of the hippocampal regions are considered  in
the functional mapping: deep layers of the medial entorhinal
cortex (denoted by EC V/VI), superficial layers (EC II and EC III)
and subfield  CA1 of the hippocampus. The tri-synaptic
path (denoted as $W_{tri}$ on Fig.~\ref{fig:struct}) involving the
Dentate Gyrus (DG) and CA3 will be  collapsed into an integrated
transformation. The potential role of the DG, CA3 as well as the
Subiculum (SUB) will be discussed in the last section. For
simplicity, all areas and subfields will be referred to as
`layers'.

As all computations described above require statistical
characterization of input ensembles, sampling and processing of
the sensory input and incremental tuning (learning) are also necessary. Input processing and learning, i.e., fine tuning of the
synaptic weights that actually filter the information, are
discussed separately.

\section*{Characterization of the input to the hippocampal region}

Let $\mathbf{x}(t)\in \mathbb{R}^n$ denote the  analog valued postrhinal input to the entorhinal
cortex at discrete time $t$ where  $n$ is the dimension of the input. In this model, we limit
ourselves to square problems, which is to say that $n$ may be considered as both the number of postrhinal neurons
and the number of entorhinal neurons of the targeted layer. Let us also assume that the input
follows the dynamics described above. The postrhinal input enters the circuitry at the superficial
layers of EC through two parallel connection systems $R_{II} \in \mathbb{R}^{n \times n}$ and
$R_{III} \in \mathbb{R}^{n \times n}$, so, we assume that the number of principal cells in each
superficial layer is equal and is also $n$.  These connection systems may only transmit cortical
input to HR, so their tuning is omitted: admitting the lack of knowledge concerning the exact nature
of the parallel postrhinal inputs, we may suppose that $R_{II}=R_{III}=I$, where $I \in
\mathbb{R}^{n \times n}$ denotes the $n \times n$ identity matrix. When the process of learning the matrices is
considered, a temporal index is shown in most cases. For better readability the time index is dropped for non-tunable matrices
and in the dynamical equations.

>From EC II/III the signals are sent to the hippocampus  through the direct, i.e., EC
III$\,\rightarrow \,$CA1, and the indirect,tri-synaptic i.e., EC II$\,\rightarrow \,$CA1 pathways
(denoted by subscripts `dir' and `tri' on the right hand side of Fig.~\ref{fig:struct},
respectively).

\section*{Detailed correspondence between the functional model and the neural layers of the HR}

The formal description has some direct consequences concerning the
potential roles of the different layers of the HR. First, it is
obvious that innovation (that is the comparison of the predicted
and actual inputs) can only be stored in a layer that not only
receives the input, but is also the target of inhibitory feedback.
Due to its widespread inhibitory network, EC II is assigned to
hold the innovation. The activity at EC II is as follows:
\begin{equation}\label{e:innovationinECII}
\mathbf{n_z}(t+1) = R_{II}\mathbf{x}(t+1) + Q_{II}\mathbf{h}(t)-K\mathbf{y}(t),
\end{equation}
where $\mathbf{y}(t)$ and $\mathbf{h}(t)$ denote the activity at EC III and EC V/VI, respectively.
(Roman subscripts of the connection matrices denote the number of targeted layers.) Connections
from EC III to EC II, denoted by $K$, are assumed to be mostly inhibitory. The reason for this
assumption is that the vast majority of the deep to superficial connections are excitatory and
mostly target principal cells in EC II \cite{vanHaeften03Morphological} and the cortical inputs are
also of excitatory nature. In turn,  $K$ is the candidate connection system  that effectively
targets the inhibitory network of EC II. Here, the role of $Q_{II}$ is to whiten the innovation,
whereas the role of $K$ is to ensure that the emerging activity pattern is indeed proportional to
the required innovation.

Equation~\eqref{e:innovationinECII} and the connectivity of the HR implies that $\mathbf{y}(t)$
should be proportional to the input and is made of two terms from bottom-up and top-down
contributions. The activity of EC III is thus the following:
\begin{equation}\label{e:ECIII}
\mathbf{y}(t) = R_{III}\mathbf{x}(t)+Q_{III}\mathbf{h}(t),
\end{equation}
where $Q_{III}$ --- in accordance with the redundancy reduction principle --- is assumed to decorrelate
the activity at the targeted layer, EC III. However, decorrelation of quantity $\mathbf{y}(t)$ may
influence (distort) the innovation in EC II. This raises some doubts, because quantity
$\mathbf{n_z}(t+1)$ might be contaminated by predictable components, or its whiteness might be
spoiled. In turn, tuning of matrix $K$ should somehow counteract both problems under the constraint
that learning is Hebbian. The solution to this threefold problem is an emerging property in our
model.

As was noted earlier, CA1 has a central location since it is targeted by both layers EC  III
and EC II via $W_{dir}(t) \in R^{n\times n}$ and $W_{tri}(t) \in R^{n\times n}$, respectively:
\begin{equation*}\label{e:ICA}
\mathbf{s}(t) = W_{dir}(t)\mathbf{y}(t),
\end{equation*}
where $\mathbf{s}(t) \in R^n$ denotes the activity of CA1, if its
driving input is projected from EC III and
\begin{equation}\label{e:ICAoninnovation}
\mathbf{e}(t) = W_{tri}(t)\mathbf{n_y}(t),
\end{equation}
where $\mathbf{e}(t) \in R^n$ denotes the activity of CA1, if its driving input is projected from
EC II. Following the proposal of \cite{Lorincz98Formingindependent,Lorincz00Twophase} and supported
by the experimental findings of ~\cite{Redish01Independenceoffiring}, independent components should
be expressed in CA1. In turn, we believe transformations $W_{dir}(t)$ and $W_{tri}(t)$ realize the
actual signal separation and provide approximate independent components. We note that according to
Eq.~\eqref{e:hidden_process_input} $\mathbf{e}(t)$ should be equal to the innovation of
$\mathbf{s}(t)$. However, unlike in the superficial layers, there are no recurrent collaterals in
CA1. This means that for properly tuned $Q_{II}$, $Q_{III}$ and $K$, the two bottom-up
transformations, i.e., $W_{tri}$ and $W_{dir}$ should become effectively identical in the absence
of recurrent collaterals.

CA1 signals may leave the loop through the subiculum or they may be sent back to
the deep layers of EC via the connection system denoted by $V \in R^{n\times n}$. (On the intriguing properties of $V$ (not modeled here), see  \cite{Naber01Reciprocal}).

In line with \cite{Lorincz02Mystery}, a central function of the deep layers of EC may be pattern
completion. However, as was already noted, forcing independence does not support pattern
completion. It is also known ?that activity patterns of the deep layers of EC are not in fact independent \cite{Sargolini06Conjunctive}. This implies that `remixing' of the components is advantageous.
Of the many possibilities, whitening seems the most straightforward transformation, as it does
not increase the number of transformations within the EC-HC circuitry. The resulting patterns may
show higher-order correlations supporting the task of pattern completion. Since the internal
predictive system is based on the intensive use of recurrent connections, only CA3 and EC V/VI may
be considered. If our assumption about the roles of the superficial layers are valid, then EC V/VI
should realize the predictive system since CA3 is not supposed to receive significant input from EC
III.

Consequently, the activity at EC V/VI can be written as:
\begin{equation}\label{e:hiddenprocess}
\mathbf{h}(t+1) = M_{h}\mathbf{h}(t)+\mathbf{n_{\mathbf{h}}}(t),
\end{equation}
where predictive system $M_h$ can propagate activity $\mathbf{h}(t)$ in time,
$\mathbf{h}(t)=V\mathbf{s}(t)$ and $\mathbf{n_{\mathbf{h}}}(t)=V\mathbf{e}(t)$.  In addition to conveying information from CA1, V is responsible
for the decorrelation of the activity patterns.
$M_{\mathbf{h}}$ is an approximation of the dynamical model underlying the observations (see
Eq.~\eqref{e:hidden_process_input}). The queuing of the arrival of the two different inputs
($\mathbf{s}(t)$ and $\mathbf{e}(t)$) requires a mechanism that can maintain activity long enough
to enable integration. Experimental findings on gradually modifiable persistent activity in EC V
\cite{Egorov02Gradedpersistentactivity} may support this proposal.

At last, the deep layers project back to EC II and EC III via $Q_{II}$ and $Q_{III}$, respectively.

\section*{Learning processes}

For different reasons, 3 connection systems are assumed to
decorrelate the activity of their targeted layer: $Q_{II}$,
$Q_{III}$, and $V$. Their tuning follows the form given in Eqs.
\eqref{e:white_x} or \eqref{e:white_n}. For example, learning of
$Q_{II}$ can be given as:
\begin{equation}\label{e:whiteQ}
\Delta Q_{II}(t+1)' \propto Q_{II}'(t)(I-\mathbf{n_z}(t)\mathbf{n_z}(t)')
\end{equation}
where $\mathbf{n_y}(t)$ is the emerging activity of the targeted layer, EC II.

To arrive at the right form of innovation, connections between EC III and EC II need to be tuned.
The learning rule of $K(t)$ is supposed to satisfy a Hebbian form, similar to
Eq.~\eqref{e:ARlearn_}
\begin{equation}\label{e:K}
\Delta K(t+1)\propto \, \mathbf{n_z}(t+1)\mathbf{y}(t)'.
\end{equation}
This is the perfect learning rule, because it minimizes cost function
\begin{equation}\label{e:costinnovationinECII}
J(K)= \frac{1}{2}\sum_t \big|R_{II}\mathbf{x}(t+1) +
Q_{II}\mathbf{h}(t)-K\mathbf{y}(t)\big|^2=\frac{1}{2}\sum_t \big|\mathbf{n_z}(t+1)\big|^2.
\end{equation}
which is the Euclidean norm of $\mathbf{n_z}$. In this expression each term is a linear transform
of $\mathbf{x}$ with different time lags. The result of the learning rule is that, apart from an
arbitrary linear transformation,
$$K\mathbf{y}(t)=Q_{II}\mathbf{h}(t)+R_{II}\mathbf{x}(t),$$ is satisfied in all instances.
This is the net result, i.e., $\mathbf{n_z}(t)$ is indeed a linear transform of innovation
$\mathbf{n_x}(t)$. Quantity $\mathbf{n_z}(t)$ will be white given the learning rule for $Q_{II}$
detailed in Eq.~\ref{e:whiteQ}. In sum, learning rule Eq.~\ref{e:K} is Hebbian and adjusts the
inhibitory contribution until $\mathbf{n_z}(t)$ becomes a linear transformation of the innovation,
subject to the constraint, that both $\mathbf{y}(t)$ and $\mathbf{n_z}(t)$ are white.

Separation takes place in both the direct and the indirect pathways, so $W_{dir}$ and $W_{tri}$
should undergo tuning similar to Eq.~\eqref{e:separation}. At this point some remarks are in order.
We expect to have two separate channels, one for the input and one for the innovation, which can
basically reverse the mixing effect of the very same mixing process (see Eq.~\eqref{e:ICA_input}).
We have also seen that both separation processes would probably end up creating approximately
independent components in the same layer (CA1). First, it is necessary to ensure that learning in
the two separation pathways converges to approximately the same solution. Second, it is necessary
to schedule the activity at CA1 to avoid interference between the patterns corresponding to the
independent components of the input or the innovation. Regarding the interaction between $W_{dir}$
and $W_{tri}$, it is intriguing that while the original problem of ICA (that is when the mixing
process and the components are unknown) is truly unsupervised, by constraining the outputs of the
tunable matrix to some prescribed outputs the learning algorithm becomes supervised. Thus, the two
matrices may become identical if one channel dominates (supervises) the other. Physiological
considerations seem to suggest a possible mechanism.

Regarding separation, in the beginning the faster direct pathway may supervise the indirect one by
providing approximately independent components in CA1. We note that there is a temporal
coordination between the firing of the neurons that send information through the direct and the
indirect paths \cite{dragoi06temporal}. It is also possible that supervising signals may reach CA1
at one phase of the theta oscillations, while the signals from the tri-synaptic pathway may reach
CA1 at the other phase. Another argument is that although place fields in CA1 begin to stabilize
early (compared to the place fields in CA3) and even without input from the tri-synaptic route,
full stabilization takes much longer. We suggest that the two routes work together. The early
stabilization results in approximate independent components if the signal from EC III is
contaminated by large temporal correlations. The task of the indirect route may be to diminish this
kind of temporal dependence and to proceed with the separation of the sources, but this is a slower
process.

Following our hypothesis, tuning of $W_{dir}(t)$ and $W_{tri}(t)$  may assume two different forms
during the course of learning:
\begin{eqnarray}\label{e:separationWdir}
\Delta W_{dir/tri}(t+1)' &\propto & W_{dir/tri}(t)'(I-\mathbf{s}(t)\, f(\mathbf{s}(t))')\\
\Delta W_{dir/tri}(t+1)' &\propto & W_{dir/tri}(t)'(I-\mathbf{e}(t)\, f(\mathbf{e}(t))').
\end{eqnarray}
where $f(\cdot)$ is an (almost) arbitrary component-wise nonlinear function.

In the formal model we have seen that all these transformations are required to provide the right
information for the internal predictive model. However, this model also needs tuning in order to match the
observed signals.

The approximation of predictive matrix $M_{\mathbf{h}}$ -- as with all predictive matrices in the
model -- can be written as follows:
\begin{equation}\label{e:hiddenARlearn}
\Delta  M_{\mathbf{h}}(t+1) \propto \,\mathbf{n}_{\mathbf{h}}(t+1)\mathbf{h}(t)'
\end{equation}
This rule trains matrix $M_{\mathbf{h}}$ to optimize prediction in with? Euclidean norm norms?. Due to the
scheduled arrival, we need to suppose that the time window is broad enough to enable interaction of
the transformed input signal and the innovation. As we see, training is Hebbian, but a detailed
mechanism that would actually be able to carry on this tuning is missing. Nevertheless, we conjecture that the
double loops of the direct and indirect pathways have a fundamental role in tunneling the right
information at the right time. It is worth noting that this assumption is also supported by the
experimental finding that activity in CA3 under one theta oscillation (50-80$\,$ms) may correspond
to 1 second of the external sensory flow. Unfortunately, available experimental data is not
sufficient to better model this interplay.

In summary, if all transformations are optimally tuned, then (1), temporal correlations $F$ are
learnt and represented in the internal model through matrix $M_{\mathbf{h}}$, (2), the hidden
processes $\mathbf{h}$ can be estimated by the learnt model and (3), the true independent causes
$\mathbf{e}$ can also be revealed. Note that two main goals are achieved; the independent causes
($\mathbf{e}$) are revealed up to an arbitrary permutation, scale and sign, and the predictive
matrix $F$ is learnt up to a linear transformation.

In the next section we turn back to the original problem of the emergence of particular spatial
activity at different parts of the HR. In the simulations the transformations assigned to different
parts of the loop are implemented and applied on structured high dimensional inputs containing
spatial information. The goal is to study whether the emergent activity at the different modules
corresponding to e.g. CA1 and ECII/ECIII resembles that found experimentally.

\section*{Results III: Simulations}\label{s:resultsIII}

We present a series of simulations with inputs of increasing
complexity. The more realistic the inputs, the more complex the
computations that are required to extract spatial information. In doing
so, the role of different modules can be highlighted.

\begin{figure}
\includegraphics[angle=-90,width=8cm]{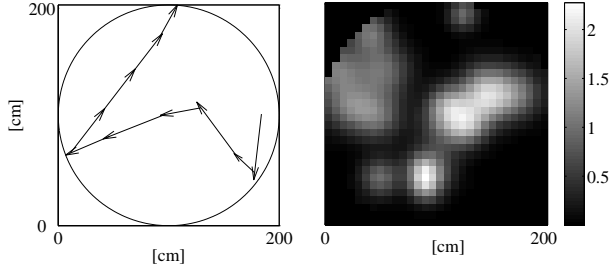}\\
\caption{(a): Circular maze, diameter: 2m, with a short sample trajectory. Step size varies between
11 and 55 cm. (b): Sample input to the loop in the form of an activity map within the maze (see,
Eq.~\eqref{e:MMG} for details). Activity map is shown in arbitrary units.}\label{fig:maze}
\end{figure}

In our sample simulations a virtual rat has explored a 2~m wide, open-field circular maze. Similar results were reached using a square maze. The path has been generated as follows: the rat runs on a
linear path at a constant speed and makes a small random turn at each step with a given chance. It also
makes a random turn if it `senses' that it may collide with the wall. Input sampling has been
fixed to 55~cm. The length of this random trajectory and input sampling were chosen to get a fair
coverage of the full area of the maze with a reasonable number of samples. The maze and a sample
trajectory is shown in Fig.~\ref{fig:maze}. Inputs corresponding to turns may only be interpreted
by higher order autoregressive processes for which the order would be about the average number of steps in a single direction. As the implemented internal model assumes first order processes (see the
comment at Eq.~\eqref{e:hidden_process_input}), such inputs have been excluded. We shall come back
to this point in Section~\ref{s:disc}.

The most restrictive approximation in our simulations is that the input contains information about
the local cues only, no distal information is included. One might think that the input is a mixture
of smells that differs from point to point. This local nature implies that parametric maze
distortions can not be modeled in this framework. On the other hand, this simplification excluded
any artifact that would result from arbitrary modeling of low-level sensory processing. Instead, we
simply mimicked postrhinal (`parahippocampal' in primates) ~\cite{Burwell03Positionalfiring}
inputs. In contrast to perirhinal input ~\cite{Eacott05Therolesofperirhinal}, postrhinal input is
assumed to reflect changes of spatial properties or directly carry spatial information (albeit in
weak correlations, ~\cite{Fyhn04Spatialrepresentation}). Such spatial dependence of the postrhinal
activity was approximated by first creating $n$ Gaussian patches with each Gaussian having a
maximum amplitude of 1:
\begin{equation}\label{e:Gauss}
g_i(\mathbf{p}) =  \exp \left( - \frac{(\mathbf{p}-\mathbf{c}_i)^2}{\sigma_i^2}\right),
\end{equation}
where $\mathbf{p}\in\mathbb{R}^2$ denotes the coordinate vector of the rat,
$\mathbf{c}_i\in\mathbb{R}^2$ is the coordinate vector of the center of the $i^{th}$ Gaussian, and
$i \in \{1, \ldots, 1000\}$. Centers $\mathbf{c}_i$ were drawn from the uniform distribution over
the full maze while $\sigma_i$ were uniformly drawn from the range  [20~cm,$\,$40~cm].

Input $\mathbf{x}$ was created by using a random, binary mixing matrix $G \in [0,1]^{1000
\times 1000}$ over the set of the Gaussians:
\begin{equation}\label{e:MMG}
\mathbf{x}(t) = G \, \mathbf{g}(\mathbf{p}(t)),
\end{equation}
where  $\mathbf{p}(t)$ denotes the coordinates of the rat in the maze at time $t$ and the $i^{th}$
component of vector $\mathbf{g} \in R^{1000}$ is $g_i(\mathbf{p}(t))$ at time $t$.  Each row of
matrix $G$ contains 20 positive non-zero elements on average. The resulting activity map for a
single component of $\mathbf{x}(t) \in R^{1000}$, i.e., for one of our `sensors' is shown in
Fig.~\ref{fig:maze}(b).

In simulation \#1 the input to the model was exactly as defined in ~\eqref{e:MMG}.

In simulation \#2 50 more units were added, so the dimension of the input, $\mathbf{x}(t)$ was 1050.
The new  units `sensed' directions and had no spatial dependence. The direction sensitivity has
been defined as:
\begin{equation}\label{e:dirsens}
x_i=f_i(\phi) = \max(0, \cos(\phi-\phi_i)),
\end{equation} for $i \in \{1001, \ldots 1050\}$,
where $\phi$ denotes the direction between the last and the current positions and $\phi_i$ denotes
the direction for which the $i^{th}$ component ($1000 < i \le 1050$) is the most sensitive. This
particular choice results in broadly tuned ($\sim \pi/2$) directional activities.

In simulation \#3, instead of mixing the units that carry different information, we used 1000
conjunctive inputs that carried spatial \emph{and} directional information:
\begin{equation}\label{e:MMGxD}
\x_i(t) =  f_i(\phi(t)) \, \left[G \,
\mathbf{g}(x(t),y(t))\right]_i
\end{equation}
where $\phi(t)$ is the direction of the rat at time $t$.

Last, in simulation \#4, we used low-pass filtered versions of the inputs of simulation \#3:
\begin{equation}\label{e:TC}
\x^{(tc)}(t+1) = (1-\alpha)\x^{(tc)}(t)+ \alpha \, \x(t+1)
\end{equation}
where superscript `$tc$' stands for `temporally convolved'. This is essentially the simplest
autoregressive process regarding Eq.~\eqref{e:hidden_process_input}.

%----------------------------
\section*{Spatial analysis}
%----------------------------

As opposed to real  spiking data, linear transformations may give rise to negative signals. In
turn, the correspondence between the unit activity values after each transformation and the
neurons' responses is not straightforward. In order to generate the activity maps of the input units,
first we discretized the space (the resolution was $30 \time 30$ so a bin is
$6.67\,$cm$\times6.67\,$cm, which is comparable to ~\cite{Hafting05Microstructure}), and for each
bin we summed up the activity measured in those steps that ended in the given bin. This spatial
averaging smoothes out the artifacts caused by unattended spots. The activity after, e.g.,
decorrelation may assume negative values, so the data were half-wave rectified (clipped) and scaled
to range $[0,1]$.

ICA is invariant for the change of sign ~\cite{Jutten91Blindseparation}. In turn, the sign of an
activity map has been defined by the average sign of the first 10 bins with the highest absolute
value. That is, if more than 5 units were negative, we simply flipped the sign of the map. The
resulting maps were then  half-wave rectified. We also computed the 2 dimensional normalized
autocorrelation for each activity map.

The spatial analysis of the peak activity regions for the autocorrelation image has been done by
fitting a grid on the locally maximal points using Delaunay-triangulation
~\cite{Markus95Inteactions,Takacs07Simpleconditions}. Border vertices and nodes have been excluded
from the analysis. Vertices are considered as internal if they belong to two triangles and nodes
are internal if they only connect to other nodes through internal vertices. To characterize the
regularity of the resulting grids, we calculated the vertex length and the angle distribution.
Discretization, however, defines a lower bound of the edge length, which is about 2 bins, that is
$\sim 13.34\,cm$. Because the mean angle in Delaunay-triangulation is obviously 60 degrees, the
spread around this value (that is the standard deviation, or std for short) can be used to quantify
regularity. The distribution of the \emph{mean} vertex lengths and the distribution of the std of
the angle values for the whole population have been used to compare the spatial characteristics of
the input set and the set of the transformed signals.

For simulations \#2, \#3, and \#4, direction sensitivity has also been analyzed. To show the
spatial distribution of the direction sensitivity, we discretized the activity maps into
$10\times10$ bins and in each bin we collected those steps that ended in that bin. Their direction,
weighted by the response value at the end point, was then added up. The resulting directed activity
values can be visualized in a `direction-field' plot. In order to characterize the spatial heterogeneity
of the directional selectivity, the directed values may also be grouped according to their
direction and these lumped sum values will be presented on a polar plot. These analysis serve to
characterize the strength of spatial heterogeneity in direction selectivity.

%----------------------------
\section*{Simulation \#1: direction-independent input}
%----------------------------

Standard PCA reduced  the dimension of the decorrelated input from 1000 to 852 as the remaining
eigenvalues were below the level of numerical precision. The first 127 dimensions carried 95\% of the total
variance and the first 203 carried 99\%. The resulting activity maps are comparable to the firing
rate maps shown in ~\cite{Hafting05Microstructure}.

%\begin{figure}[t!]
%  \includegraphics[width=12cm]{MMG_PCA_grid_examples_7_21_65_89_and_stat.eps}\\
%  \caption{Position dependent input. (a-d): each column corresponds to different output units.
%  First row: activity maps with negative values clipped and then scaled to interval [0,1] with 0: black, 1: white.
%  Second row: 2D autocorrelation function of the activity maps and the fitted grids.
%  Third row: Vertex angle histogram for the fitted grids. Cumulative statistics over all grids.
%  (e): histogram of the mean edge length of the grids for the input set and the PCA units.
%  (f): histogram of the standard deviation of the vertex angles for the input set and the PCA units.
%}\label{f:grids}
%\end{figure}

\begin{figure}[t!]
\includegraphics[width=10cm]{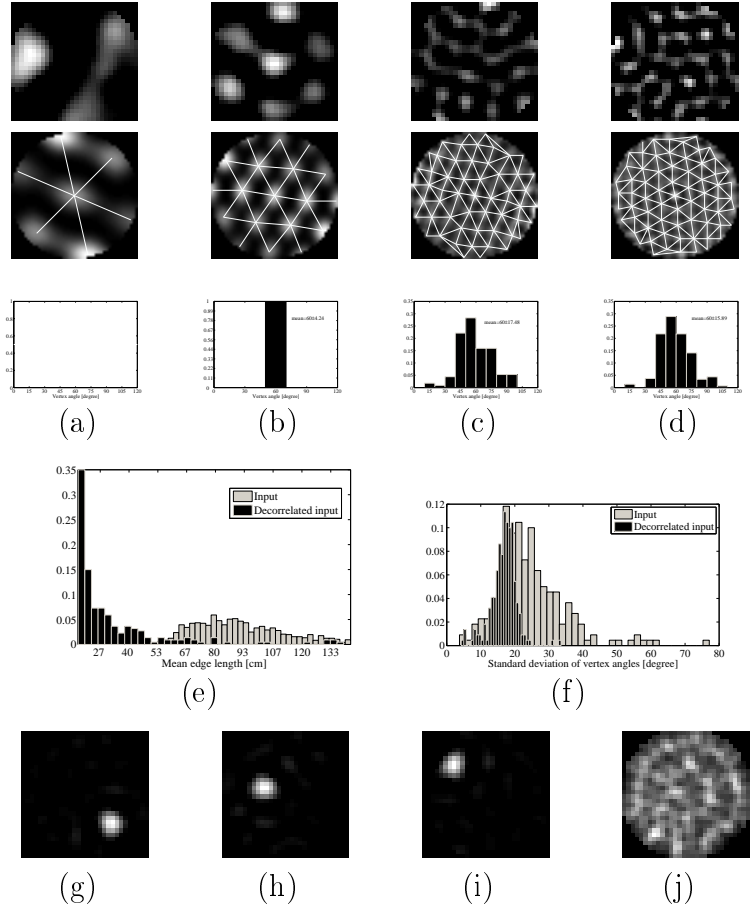}\\
\caption{Simulation \#1: position dependent input. (a-d): each column shows the output of different
\emph{decorrelating} units (PCA). First row: half-wave rectified and scaled activity maps (0:
black, 1: white). Second row: 2D autocorrelation function of the activity maps and the fitted
grids. Third row: vertex angle histogram for the fitted grids. (e-f): cumulative statistics over
all grids. (e): histogram of the mean edge length of the grids for the input set and the PCA units.
(f): histogram of the standard deviation of the vertex angles for the input set and the PCA units.
(g-i): sign corrected activity maps of three \emph{separating} (ICA) units. Response is localized.
(j): superimposed map of all ICA units demonstrating that the localized units cover the full
maze.}\label{f:MMG_PCA_ICA}
\end{figure}

Subfigures~\ref{f:MMG_PCA_ICA} (a-d) show the spatial activity of 4 sample units. The first row
depicts the clipped activity maps. We put the 2D autocorrelation functions of the activity maps and
the superimposed grids into the second row. Note that the peak-to-peak distance varies over a broad
range. The third row represents the vertex angle distribution of the corresponding grids. Narrower
distribution means more uniform vertices and thus a more symmetric grid.
Subfigures~\ref{f:MMG_PCA_ICA}(e) and (f) show cumulative statistics concerning the grids. Only the
grids of the first 220 largest eigenvalues have been used in these analyses as the rest are mostly
noise. Subfigure~\ref{f:MMG_PCA_ICA}(e) compares the distribution of the mean vertex length of the
fitted grids for the input activity maps and the activity maps of the decorrelated inputs.
Subfigure~\ref{f:MMG_PCA_ICA}(f) compares the distribution of the standard deviation of the vertex
angles of the fitted grids. Again, for hexagonal-like grids, the smaller the standard deviation,
the larger the regularity. While the experimental data available to us is not sufficient for
comparisons, we can safely claim that these grids do cover a range similar to those of
\cite{Hafting05Microstructure}. These diagrams unambiguously show that the `gridness' of activity
has increased significantly due to the decorrelation step. Since the creation of the input is
essentially equal to a random mixture, the effect we show here is not an artifact.

For the sake of completeness, the effect of separation is also shown. It was demonstrated in
\cite{Takacs07Simpleconditions,Franzius07Fromgrids} that inputs with grid-like spatial activity
patterns can be transformed into more localized `place cell'-like activity patterns by imposing
independence or sparseness on the components of activity. As decorrelation in our simulations
already yields grid activity, separation (into independent components) naturally resulted in
unimodal place-cell like activity maps. Subfigures~\ref{f:MMG_PCA_ICA}(g-i)  show three sample
units and subfigure~\ref{f:MMG_PCA_ICA}(j) depicts the superposition of activity maps of 60
independent units. The resulting coverage may be interpreted as coarse grain disretization of a
low-dimensional space (in our particular case, the relevant dimension is 2).

%----------------------------
\section*{Simulation \#2: Mixture of direction-independent and position-independent inputs}
%----------------------------

\begin{figure}
\includegraphics[width=8cm]{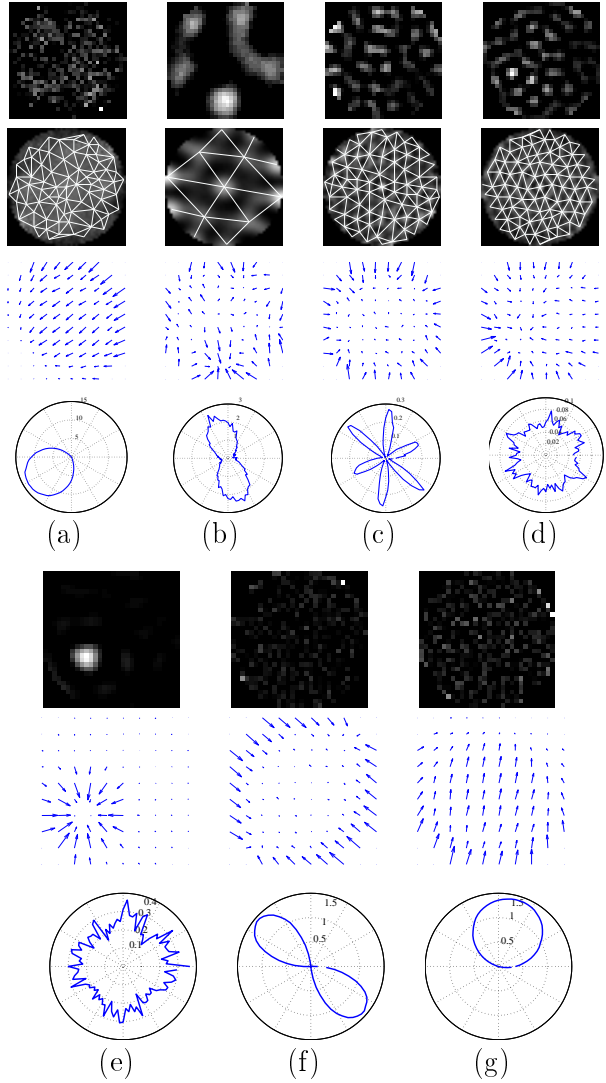}
\caption{Simulation \#2: mixture of purely position and purely direction selective inputs. (a-d):
columns correspond to the output of different \emph{decorrelating} units (PCA). First row:
half-wave rectified and scaled activity maps. Second row: 2D autocorrelation function of the
activity map and the fitted grid. Third row: spatial distribution of the direction selectivity is
shown on a square grid of size 10x10. Fourth row: overall direction selectivity in the form of a
polar plot. (e-g): columns correspond to the output of different \emph{separating} units (ICA).
First row: sign flipped, half-wave rectified activity maps.  Second row: spatial distribution of
the direction selectivity is shown on a square grid of size 10x10. Third row: overall direction
selectivity in the form of a polar plot.}\label{f:MMG_D_PCA_ICA}
\end{figure}

After decorrelation most grid structures were almost identical to those of simulation \#1
(Fig.~\ref{f:MMG_D_PCA_ICA}(a-d)). However, many units also showed a certain degree of direction
selectivity, too. Clearly this ensemble of units with different dependencies shows an apparent
conjunctive representation of position and direction. Similar representation was found in dMEC
III-V ~\cite{Sargolini06Conjunctive}. However, separation after decorrelation unambiguously shows
(Fig.~\ref{f:MMG_D_PCA_ICA}(e-g)) that there are now 2 relevant subspaces: direction and position.
While most cells showed place-cell like activity (e.g. Fig.~\ref{f:MMG_D_PCA_ICA}(e)), the rest of
the units showed no spatial dependence: they were selective only for direction. For each subspace,
separation essentially resulted in a coarse grain discretization. We emphasize that the decoupling
of the directional information did not require any predictive mechanism in this case. It can be
explained by the fact that a subgroup of the original inputs contained explicit directional
information so their activity statistics are obviously different from the other units in the linear
mixture.

%----------------------------
\section*{Simulation \#3: Position and direction dependent inputs}
%----------------------------

\begin{figure}
\includegraphics[width=8cm]{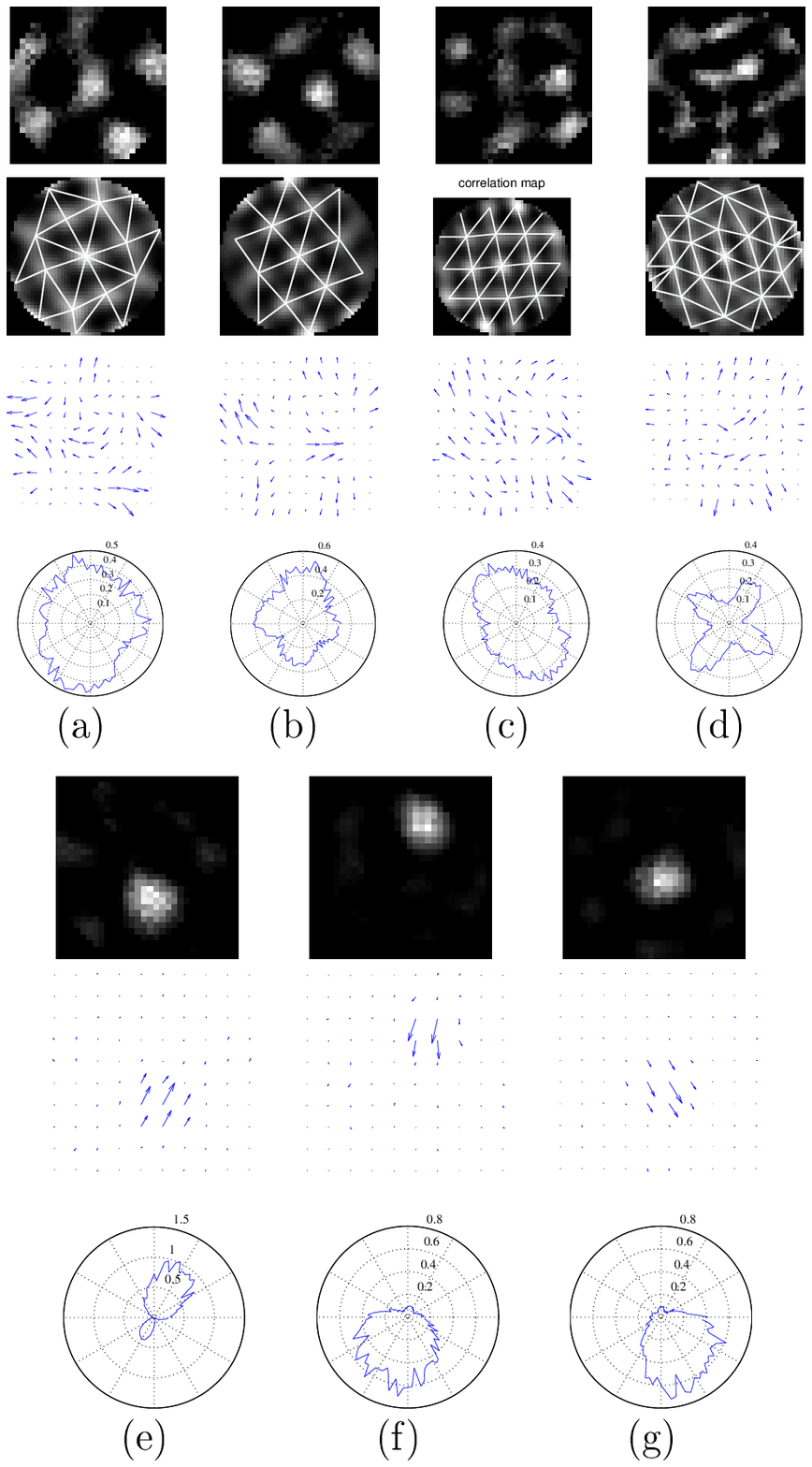}
\caption{Simulation \#3: position \emph{and} direction selective inputs. (a-d): columns correspond
to the output of different \emph{decorrelating} units (PCA). First row: half-wave rectified and
scaled activity maps.  Second row: 2D autocorrelation function of the activity map and the fitted
grid. Third row: spatial distribution of the direction selectivity is shown on a square grid of
size 10x10. Fourth row: overall direction selectivity in the form of a polar plot. (e-g): columns
correspond to the output of different \emph{separating} units (ICA).  Fifth row: sign flipped,
half-wave rectified activity maps. Sixth row: spatial distribution of the direction selectivity is
shown on a square grid of size 10x10. Seventh row: overall direction selectivity in the form of a
polar plot.}\label{f:MMGxD_PCA_ICA}
\end{figure}

For true conjunctive inputs (i.e. all input units show both position  and direction selectivity),
two changes can be seen in the activity maps of the decorrelating units
(Fig.~\ref{f:MMGxD_PCA_ICA})(a-d). First, all units inherited the conjunctive property showing some
direction selectivity on top of the grid-like spacing. Second, regularity and symmetry properties
degraded in both subspaces compared to those of Fig.~\ref{f:MMG_D_PCA_ICA}(a-d). The output of the
separating units (ICA), in contrast to Simulation \#2, now all showed significant direction
selectivity as well (Fig.~\ref{f:MMGxD_PCA_ICA})(e-f).  It implies that the ICA units are again
\emph{local}, but now in 3 dimensions: ICA cells basically discretize the Cartesian product of the
2 dimensional maze and the 1 dimensional space of directions.

\begin{figure}
\includegraphics[width=6cm]{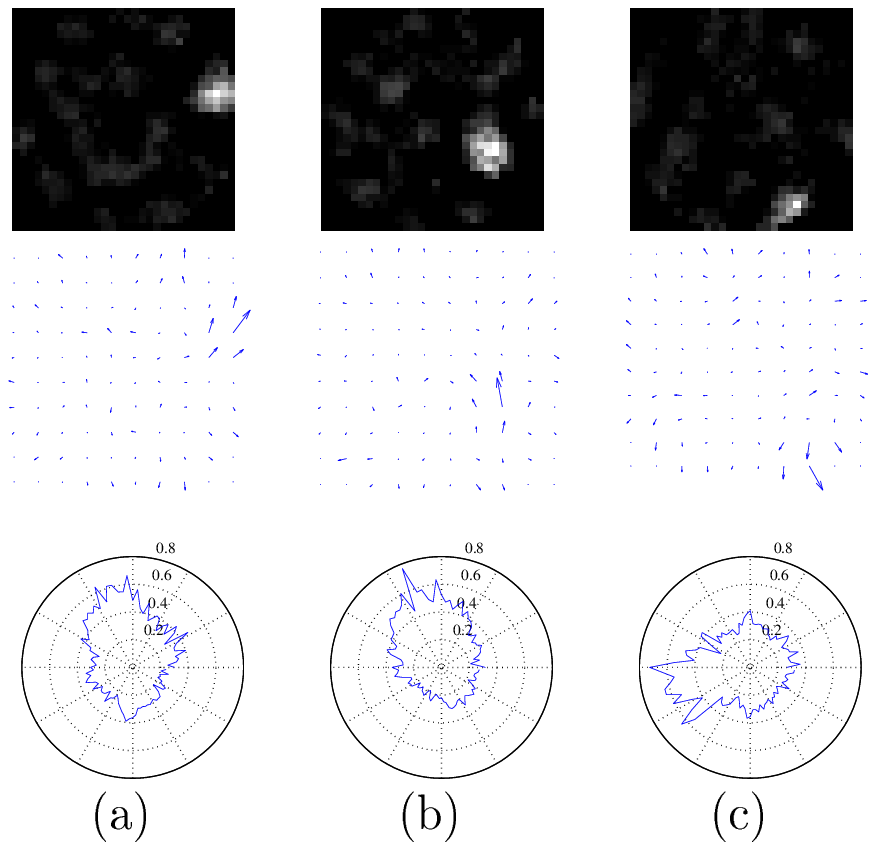}
\caption{Innovation on the decorrelated conjunctive inputs. (a-c): columns correspond to the output
of different separating (ICA) units. First row: half-wave rectified activity maps.  Second row:
spatial distribution of the direction selectivity is shown on a square grid of size 10x10. Third
row: overall direction selectivity in the form of a polar plot.}\label{f:MMGxD_ICA_on_innovation}
\end{figure}

According to our hypothesis, the internal model realized in the deep layers of the entorhinal
cortex is responsible for restoring the decoupling between predictable (i.e., direction selective) and
nonpredictable information. In turn, we expect to see weaker direction selectivity if the
predictive model is also part of the computations (the circuitry now works on the difference
(innovation) between the input and the expectation of the system's internal model, see
Eq.~\eqref{e:innovationinECII}). If the internal model is correctly tuned, then  directional
sensitivity should disappear from the innovation, resulting in clear hexagonal spacing again.
Furthermore, we should see a diminished direction selectivity in place cell activity as well.
Indeed, the directional sensitivity of the innovation of the decorrelated inputs decreased
considerably, although the hexagonal structure of Fig.~\ref{f:MMGxD_PCA_ICA} did not improve
significantly (results are not shown). In addition, separation of the innovation yielded local
activity with diminished direction selectivity as predicted (Fig.~\ref{f:MMGxD_ICA_on_innovation}).

%----------------------------
\section*{Simulation \#4: Temporally convolved position and direction dependent inputs}
%----------------------------

\begin{figure}
\includegraphics[width=6cm]{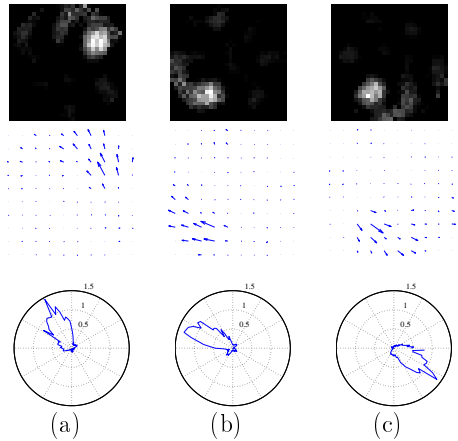}
\caption{Separation on temporally convolved, position and direction dependent inputs. (a-c): each
column corresponds to different ICA output units. First row: sign flipped and half-wave rectified
activity maps. Second row: spatial distribution of the direction selectivity is shown on a square
grid of size 10x10. Third row: overall direction selectivity in the form of a polar
plot.}\label{f:TCMMGxD_ICA}
\end{figure}

As we noted earlier, theta oscillation may be responsible for time compression in the HR, which --
from the computational point of view -- corresponds to temporally convolved inputs. The resulting
moving average would probably highlight those factors that change less abruptly. In our case,
apart from the turns taken in order to avoid collisions, directional information is either constant or varies
slowly over a longer time interval. In turn, we expect to see that separation on the temporally
convolved inputs collected during linear motion would yield stronger direction sensitivity with
larger and less precise place fields (Fig.~\ref{f:TCMMGxD_ICA}), similar to those found in the
different areas of the subicular complex ~\cite{Sharp96Multiplespatial}. Direction selectivity has
a much \emph{finer scale} than the half-rectified cosine dependence used for the inputs.

\section{Discussion}\label{s:disc}

In this last section, we analyze the simulation results and re-evaluate the functional mapping of
our computational model. These considerations then lead to some predictions about the HR. We conclude with a discussion of a few issues still left unresolved and identify possible further improvements.

\section*{Interpretation of the simulation results}

Although our model construct is based on general ideas about efficient representation of
sensory events, when applied to spatially anchored inputs it has shown some intriguing
properties that can directly correspond to experimental data.

The model correspondences have already been supported by the first simulation in that grid-like
activity has appeared in exactly those modules the neural substrates of which were reported to present
this particular activity. Once grid-like activity is present, forcing independence results in
localized activity, as was shown for example, in \cite{Franzius07Fromgrids}. What is more
interesting, though, is that reciprocity (i.e., place cells are needed to get stable grid cells) can
also be explained by the loopy structure of our model. Another observation is that the weak overlap
among the resulting place fields can be considered as \emph{discretization} of the space. Similarly
to what was found first in \cite{Lorincz01Independentcomponent}, this is what ICA seems to do if
there is a small dimensional space behind the high dimensional inputs.

In Simulation \#2, directional information was introduced by mixing output of position
dependent and purely direction dependent units. Let us remark here that no such labeling as
position or direction was given in the model. The proposed algorithms simply extract
statistical properties of the input ensemble. This simulation yielded two interesting results. The
first one is that grid cells now show \emph{conjunctive} behavior as well: in addition to pure position
dependence, many units appear to depend on both space and direction. The other result that
independent components now either show spatially localized activity with no direction selectivity
or demonstrate clear direction-selectivity without any particular position dependence is a
consequence of our input preparation, because we mixed two signals with different statistical
properties and ICA separated them. Now, ICA discretizes \emph{two} distinct subspaces separately;
the space of position and the directional space. This is not unlike how information flow is
supposed to take place between CA1 and subiculum. It is, however, interesting that despite the
significant temporal correlation carried by directional information, separation alone (i.e. without
activation of the predictive system) was able to decouple these different pieces of information.
These observations also show the robustness of the ICA algorithm.

In Simulation \#3, all input units had position and direction dependence (sampled from the
product space of positional and directional information). In this case decorrelation and extraction
of the components yielded distorted, less regular grid activity and all independent components showed directional dependence as well. Now, separation discretized the 3 dimensional conjunctive space of position and direction. However, turning on the predictive system significantly lessened the directional selectivity of the EC II grid units, in accordance with the experimental findings. In turn, the extracted independent components of the EC II outputs
also showed less directional dependence.

In Simulation \#4 we were interested in what happens if correlations at \emph{different} time
scales (directional information and temporal convolution) are present in the input. The most
important result here is that convolution (`moving-average') actually facilitates the decoupling:
narrowly tuned direction selective units appeared with large, less precisely defined place fields.
This kind of activity is similar to what has been reported on the subicular complex.

\section*{Consequences and predictions}

The results presented here have some direct consequences for the mapping. As decorrelation clearly yields grid-like activity and such activity has been found in the deep EC layers as well, arbitrariness in the choice of the CA1 afferents of EC deep layer (matrix $V$) may be resolved.
Decorrelation seems appropriate. Another consequence is that depending on the temporal structure of
the input, after realizing these particular correlations separating transformations may efficiently channel the information. Although we omitted the modeling of the subicular complex, this observation may explain the existence of the distal/proximal loops between CA1, subiculum and EC ~\cite{Gigg06Constraintsonhippocampalprocessing}.

Before presenting our conjectures and predictions, let us recap the logic behind them. First we claimed that a memory system is efficient if the resulting representations (1) support a predictive internal model of sensory events, (2) can be interpreted in a probabilistic framework to cope with uncertainties and (3) can be factored to maintain the redundancy reduction principle, but also help reveal relevant subspaces. These high level functional motivations lead to
a computational model that can explain the sensory input in terms of independent causes and can also predict the temporal changes of these causes and their interactions. The predictive faculty of the proposed structure is realized in an internal model that can take into account intrinsic (e.g. self-motion induced) and extrinsic changes in the observed signals. It is worth noting here that such distinctions are only meaningful if \emph{control} of the intrinsic changes (for instance, changing the pace through appropriate motor commands) is possible.  The required computational stages form a loop in which learning (tuning) and functioning are tightly coupled. The loopy structure implies that the HR connects the downstream and upstream information flow between the efferent and afferent pathways.

Next we attempted to map the proposed functional model onto the neuronal substrate by enumerating supporting anatomical, physiological and behavioral data.  Due to complexity of the problem a series of simplifications had to be introduced. Our large scale functional model ignores fine temporal scales, thus (1) rate based coding of information is sufficient. We also reduce the difficulty by focusing only on (2) linear transformations -- apart from the rectification of the neuronal outputs -- although each stage can also be extended to be nonlinear. We intend to provide
(3) a network level description only, in which the transformations are carried out by similar computational units. These considerations together with the validating simulations, which were specifically aimed at studying spatial dependence,  may lead to the following conjectures:

\begin{enumerate}
\item  The core transformation of the circuitry may be seen as a realization of independent process analysis which provides a two stage solution to recover hidden components as well as the dynamics.
\begin{enumerate}
\item In one stage separation of independent (hidden) causes and
their corresponding subspaces may take place. The HC plays a crucial role in mapping independent coordinates such as position and direction to different areas. Grouping of the components of the non-independent factors may occur by using the information about their `non-independence', i.e., within the subspaces themselves.
\item   In the other stage, a predictive system is implemented that can be fine tuned to fit the temporal scale of the evolution of the observed signals. Due to the interplay between these two functions, decorrelation and separation take place repeatedly along both the direct and the tri-synaptic routes.
\end{enumerate}
\item Depending on the capacity of the available resources, separation can be seen as a means
(1) of finding separable \emph{subspaces} and (2) of \emph{discretizing} these low dimensional subspaces. Position and direction, for instance, can be seen as two complementary but independent pieces of information. In turn, separation has a central role in shaping the responses of both the place and the head direction cells.
\item The predictive internal model is maintained by EC V/VI
\item The innovation term is formed in EC II through a complex interaction of at least 3 different areas projecting to the given layer.
\item The main input is held in EC III
\item The innovation term is the net result of the comparison of the expected input produced by the predictive system and the real input. Such comparison is made possible through the activation (by the EC III to EC II connections) of the widespread inhibitory network of EC II.
\item For both the innovation and the input, bottom-up and top-down connections work in concert to achieve decorrelation.
\item Actual separation is carried out in both the direct and the tri-synaptic pathways, resulting in independent activity in CA1. The two processes interact during learning as well as functioning.
\item Forcing independence may interfere with prediction, so some remixing is needed. Whitening seemed to be a natural choice, and this was supported by the comparison of the simulation results (grid activity emerges by decorrelation) with the experimental findings (grid activity can be found in all layers of dMEC).
\item The loopy structure and the whitening role of EC deep to EC superficial connections explains the fact that when the HC is removed signals of both superficial layers of EC change \cite{Fyhn04Spatialrepresentation}.
\end{enumerate}

The resulting mapping is an improvement over the one proposed in
\cite{Lorincz00Twophase,Lorincz02Mystery} where decorrelation was assigned to CA3. This modification is necessary \cite{Takacs07Simpleconditions}, because in applying decorrelation to spatially defined inputs grid like activity emerges and such grids were found in the entorhinal cortex and not in the CA3. In our model the main role of the deep-to-superficial connections is whitening, whereas the comparator role of EC layer II \cite{Lorincz00Twophase,Lorincz02Mystery} has not been modified.
%This finding also provides a constraint on the nature of the
%CA1->Subiculum->EC V/VI connections.

Regarding spatial information, simulations revealed that decoupling of directional and positional
information is viable in our model framework. If the neuronal mapping is correct, this decoupling
defines the interplay between the hippocampus proper (responsible for shaping and maintaining
primarily positional information) and the subicular complex (responsible for directional information).
It may also explain the necessity of the two parallel routes. Because the proximal and distant
targets differ in the two areas, it is possible that computations are similar at the CA1 and at the
subiculum.

These considerations imply the following predictions:

\begin{description}
\item[Subspace separation] At the initial stage of place field stabilization in CA1 cells may show gradually diminishing direction sensitivity. If this conjecture is not supported by experimental findings, then place field formation cannot be explained by applying purely statistical considerations.
\item[Top-down influences] The key role of the deep layers of EC in extracting temporal dependencies (i.e. separating the predictable parts) implies that perturbation at these layers would result in a faulty prediction system and a weaker representation of directions in the subicular complex.
In particular, changes in the activity of EC superficial layer neurons are expected. If such changes indeed exist, then the characteristic properties of these changes provide information about top-down influences on input filtering: modulation of the internal dynamical model may change the information that traverses to the CA1 subfield.
\item[Distortions] In the model, parametric distortion of the grids
(\cite{Barry07Experiencedependent}) may only be demonstrated by providing information about the motor efferents or by providing access to control the processing of sensory information.
\end{description}

Processing along the direct pathway is faster, as fewer transformations are involved. However, when temporal correlations are present, the resulting components may be distorted. In this case, the tri-synatic pathway is expected to become dominant, as can diminish these correlations. In sum, the varying influence of the two pathways may cause the temporary direction selectivity of the emerging place fields.

In our proposal, learning in the direct and tri-synaptic pathways takes place at different speeds.  As independent sources should be developed on i.i.d. sources we would expect that the CA1 responses are defined by the tri-synaptic pathway at least at the fine tuning stage of learning. According to the experiments
\cite{Sybirska00prominence}, each pathway can form stable place
fields in the absence of the other. The processing along the
direct pathway is probably faster \cite{leutgeb04distinct} and we
think that this is due to lack of temporal decorrelation in this
pathway. However, when temporal decorrelation is present, the
tri-synaptic route may take the lead in tuning CA1. Both proximal
and distant dendrites may play a role in learning the separation
transformations, especially in the coordination of the ICA
components.

The second prediction emphasizes the fact that physical constraints of the animal's motion set the
temporal scale of changes in direction. If the predictive internal model cannot correctly register
this timescale, then extraction of the this kind of information will be impaired while recovery of
positional information remains intact.

The last prediction deserves some comments. As the main computations in our model are aimed at characterizing a \emph{set} of inputs by extracting statistical information, any change in the
underlying statistics would result in strong distortions of the emerging activity pattern (see
Simulations \# 3 and \#4). Introducing control would affect the expectations of the internal model, which in turn would modify the predictions as well. If control information can be used as an internal metric, then it may help to recalibrate allocentric sensory information by modifying the expectations. This process, however, differs from relearning the underlying statistics. Although a surprisingly large number of properties can already be shown by our simple model without the context of any information on motor actions, it is known \cite{buzsaki_book} that maps in the hippocampus need motor actions and also dead reckoning builds on explorative trajectories with long and almost straight segments with intermittent random turns.

\section*{Falsifying issues}

We consider the consequences of our predictions crucial in our model concept. The first issue is that our model relies heavily on the strong coupling between bottom-up and top-down information for both the whitening and the separation stages. In addition claim concerning basically identical transformation (the similarity between $W_{dir}$ and $W_{tri}$) is quite restrictive. If this
constraint is not experimentally supported then serious
reconsideration seems necessary.

The other issue regards the effect of goal-oriented behavior.
Although we have seen that our model yields orthogonal hexagrid
tiling, the resulting grids are not oriented. \emph{Oriented}
grids, however, may not be formed without additional constraints
in our model. Based on the arguments concerning the differentiation of
internal and external observation signals we believe that integration with control \cite{szita04kalman} over the observation process could yield the desired property.

\section*{Open issues}

While the model we have proposed successfully replicated the reported space-dependent activity at different areas of the HR, several questions remain unanswered. First, we enumerate issues related either to the current stage of our model construction or to the particular form of the presented simulations.

In our simulations we used locally defined inputs and did not model sensory associations between
local and distal cues. Such binding is not trivial and remains a hot issue for example in computer
vision.

Although we showed that separation  of relevant low-dimensional subspaces is possible, the
mechanism of regrouping or \emph{fusion} of the factors belonging to the same subspace is not yet
known. We suppose that the particular cross talk between CA1 and the subiculum
\cite{Gigg06Constraintsonhippocampalprocessing} may provide a clue.

As regards prediction, even for the simplest case of the first order autoregressive process the
training of the predictive matrix is quite involved, as the required innovation and signal terms are
supposedly stored in different areas. In turn, queuing their arrival is very fundamental. At
present it is not known what kind of network mechanism may set the timing. As was suggested in
\cite{dragoi06temporal}, one candidate would be the network level theta-oscillation that may gate
information transfer to the deep layers of EC.  In favor of this proposal, it is known that deep
layer principal cells have distinctive theta modulation properties (see, e.g.,
\cite{Chrobak00Physiologicalpatterns} and references therein) and LTP in the deep layers of the EC
may be preferentially responsive to slow patterned activity \cite{yun02variation}.

In the following, we name a few important properties of the hippocampal region not yet integrated into the model.

One relevant question concerns how the memory system can store
information after one encounter (`one-shot' learning). This
phenomena probably requires an additional mechanism not yet
incorporated into our model since it is not based on statistical
learning principles. Such a mechanism could be simple and Hebbian
\cite{kormendy99winner-take-all}.

Setting aside this prompt learning, consolidating the acquired knowledge usually takes more time.
Presumably sequential replay of previously formed activity patterns in CA3 may facilitate this
process. In line with our initial assumptions we conjecture that forward replay may actually help shape the predictive system, while reverse replay is required to form better strategies for
goal-oriented behavior \cite{Sutton98Reinforcement}. To define goals and behavior for our system,
first a control mechanism should be integrated. Such a mechanism would affect the sampling of the
available inputs by changing the trajectory. In the simulations, we introduced one form of
temporal convolution, but it is known that HR is able to represent sequences of spatiotemporal
activity patterns in a temporally compressed form of varying timescales. Such highly versatile
convolution makes decoding even harder. It was suggested \cite{Lorincz00Twophase} that this task is assigned to the
EC-DG-CA3 loop. A further improvement of our model would be to incorporate
this loop as well.

\section{Acknowledgments}
G. Sz. is supported by the Zoltán Magyary fellowship of the
Hungarian Ministry of Education. We are grateful to Zolt\'an
Szab\'o, for his help in running some of the computer experiments.

This material is based upon work supported partially by the EC FET
grant, the `New Ties project' and EC NEST grant, the `Percept
project' under under contracts No.~003752 and No.~043261,
respectively. Any opinions, findings and conclusions or
recommendations expressed in this material are those of the
authors and do not necessarily reflect the views of the EC, or
other members of the EC New Ties or Percept projects.

% ----------------------------------------------------------------

%\clearpage\newpage
%\bibliographystyle{apalike}
%\bibliographystyle{amsplain}
%\bibliography{coexistence}

\end{document}